\documentclass[a4paper,fleqn,usenatbib]{mnras}

\usepackage[T1]{fontenc}
\usepackage{ae,aecompl}

\usepackage{graphicx}	
\usepackage{amsmath}	
\usepackage{amssymb}	
\usepackage{txfonts}

\usepackage{hyperref}	
\hypersetup{colorlinks=true,linkcolor=blue,citecolor=blue,filecolor=blue,urlcolor=blue}

\newcommand{\bea}{\begin{eqnarray}}
\newcommand{\eea}{\end{eqnarray}}
\newcommand{\bc}{\begin{center}}
\newcommand{\ec}{\end{center}}

\newcommand{\vect}[1]{\boldsymbol{#1}}

\title[Hydrogen Reionization in Illustris]{Hydrogen Reionization in the Illustris universe}
\author[A. Bauer, et.~al.]
{\parbox{17cm}{Andreas Bauer$^1$\thanks{E-mail: andreas.bauer@h-its.org}, Volker~Springel$^{1,2}$,
Mark Vogelsberger$^{3}$, 
Shy Genel$^{4}$\thanks{Hubble Fellow},
Paul Torrey$^{3}$,
Debora Sijacki$^{5}$,
Dylan Nelson$^{6}$ \&
Lars Hernquist$^{6}$}\vspace*{0.2cm}\\
  $^1$Heidelberger Institut f\"{u}r Theoretische Studien,
  Schloss-Wolfsbrunnenweg 35, 69118 Heidelberg, Germany\\
  $^2$Zentrum f\"ur Astronomie der Universit\"at Heidelberg,
  Astronomisches Recheninstitut, M\"{o}nchhofstr. 12-14, 69120
  Heidelberg, Germany\\
  $^3$Department of Physics, Kavli Institute for Astrophysics and Space Research, Massachusetts Institute of Technology, Cambridge, MA 02139, USA\\
  $^4$Department of Astronomy, Columbia University, 550 West 120th
  Street, New York, NY 10027, USA\\
  $^5$Institute of Astronomy and Kavli Institute for Cosmology, University of Cambridge, Madingley Road, Cambridge CB3 0HA, UK\\
  $^6$Harvard-Smithsonian Center for Astrophysics, 60 Garden Street,
  Cambridge, MA 02138, USA
}

\date{Accepted XXX. Received YYY; in original form ZZZ}

\pubyear{2015}

\begin{document}
\label{firstpage}
\pagerange{\pageref{firstpage}--\pageref{lastpage}}
\maketitle

\begin{abstract} 
Hydrodynamical simulations of galaxy formation such as the Illustris
simulations have progressed to a state where they approximately reproduce
the observed stellar mass function from high to low redshift. This in
principle allows self-consistent models of reionization that exploit the
accurate representation of the diffuse gas distribution together with the
realistic growth of galaxies provided by these simulations, within a
representative cosmological volume. In this work, we apply and compare two
radiative transfer algorithms implemented in a GPU-accelerated code to the
$106.5\,{\rm Mpc}$ wide volume of Illustris in postprocessing in order to
investigate the reionization transition predicted by this model. We find
that the first generation of galaxies formed by Illustris is just about
able to reionize the universe by redshift $z\sim 7$, provided quite
optimistic assumptions about the escape fraction and the resolution
limitations are made. Our most optimistic model finds an optical depth of
$\tau\simeq 0.065$, which is in very good agreement with recent Planck
2015 determinations. Furthermore, we show that moment-based approaches for
radiative transfer with the M1 closure give broadly consistent results
with our angular-resolved radiative transfer scheme. In our favoured
fiducial model, 20\% of the hydrogen is reionized by redshift $z=9.20$, and
this rapidly climbs to 80\% by redshift $z=6.92$. It then takes until
$z=6.24$ before 99\% of the hydrogen is ionized. On average, reionization
proceeds `inside-out' in our models, with a size distribution of reionized
bubbles that progressively features regions of ever larger size while the
abundance of small bubbles stays fairly constant.

\end{abstract}

\begin{keywords}
radiative transfer -- methods: numerical -- H II regions --
galaxies: high-redshift -- intergalactic medium  -- 
dark ages, reionization, first stars
\end{keywords}

\section{Introduction}
\begin{figure*}
\begin{center}
\setlength{\unitlength}{1cm}
\hspace*{-0.2cm}\resizebox{10.0cm}{!}{\includegraphics{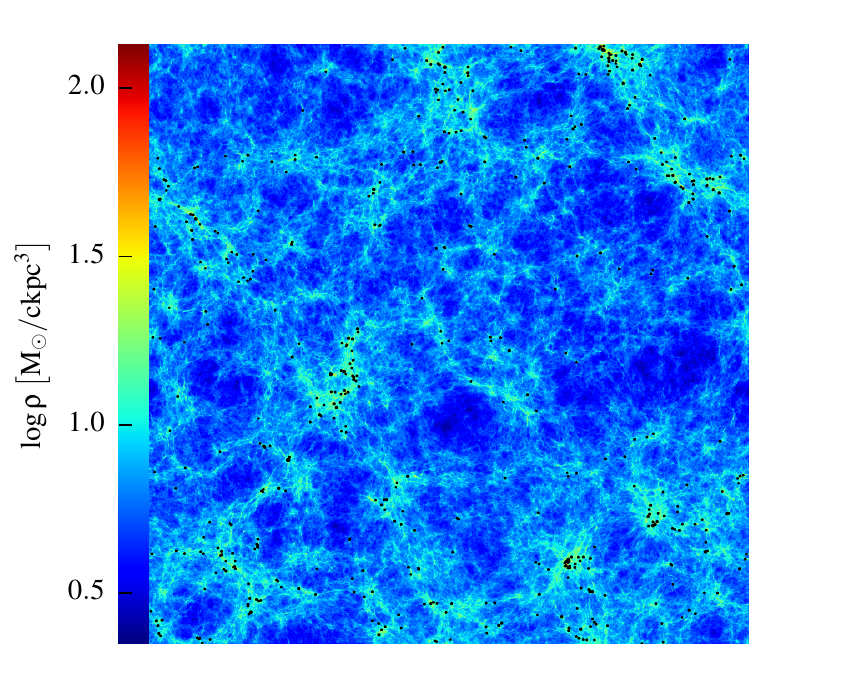}}%
\hspace*{-1.5cm}\resizebox{10.0cm}{!}{\includegraphics{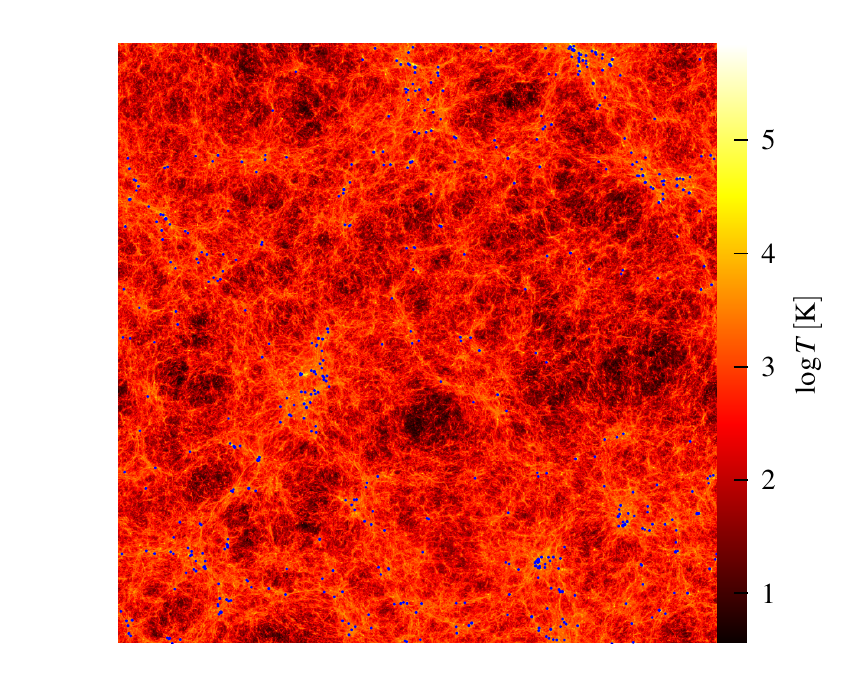}}
\caption{Projected slices of thickness $7.1\,{\rm cMpc}$ through the
  binned gas and temperature fields of the $106.5\,\mathrm{cMpc}$ wide
  Illustris simulation at redshift $z=7$, just before the externally
  applied UV background raises the temperature of the diffuse gas (the
  unit `cMpc' stands for comoving Mpc). The left-hand panel shows the gas
  density field, with overlaid circles giving the locations of
  galaxies identified at this time by the group finder algorithm
  {\small SUBFIND} \citep{Springel2001}. The right-hand panel gives the
  corresponding mass-weighted temperature field; the heated regions
  around the galaxies are caused by shocks associated with 
  virialization and feedback-driven outflows. In turn, the
  high temperature there leads to collisional ionization of the
  gas. \label{fig:slices}} \end{center}
\end{figure*}

The cosmic microwave background (CMB) radiation was released when the
Universe recombined at a redshift of $z\sim 1100$, leaving behind only
a tiny residual free electron fraction. Yet in the present Universe,
it is well established that the intergalactic medium (IGM) is highly
ionized, as is inferred from the absence of \citet{Gunn1965} troughs
in the absorption spectra of nearby quasars. Hence there must have
been an `epoch of reionization' (EoR) sometime in between, where
photons emitted by stars and possibly quasars ionized the
intergalactic hydrogen again \citep[for reviews see][]{Barkana2001,
  Fan2006, Morales2010}. This is believed to first happen to hydrogen
at $z \sim 7-10$, with helium being ionized considerably later at
$z\sim 3$  \citep{Schaye2000, McQuinn2009} by the harder radiation of quasars.

Observations of distant quasars show unambiguously that reionization
was complete not much later than $z=6$ \citep{Fan2006a}. Observations
of Lyman-$\alpha$ emitters indicate rapid changes in their abundance
at somewhat higher redshift \citep[e.g.][]{Ouchi2010, Kashikawa2011,
  Caruana2014}. Together with the relatively high optical depth to
scattering on free electrons inferred from CMB observations
\citep{Bennett2013, Planck2013CosmoParams} and the discovery of early
galaxies at $z\sim7$ and higher \citep{Bouwens2011, Pentericci2011,
  Oesch2014}, this suggests that reionization likely started
considerably earlier than $z \sim 7$. The duration of the transition
process, and the nature of the source population ultimately
responsible for reionization, remain however subject of much
observational and theoretical research. 

A particularly exciting prospect is that an observational breakthrough
in this field may be imminent, in particular through a direct mapping
of the EoR with 21-cm observations \citep[see][for a recent
review]{Zaroubi2013}. This has not yet been achieved, but impressive
progress towards this goal has recently been made
\citep[e.g.][]{Dillon2014, Parsons2014}, and future instruments such
as the Square Kilometer Array (SKA) or the James Webb Space Telescope
(JWST) promise to revolutionize our understanding of the early
universe, and of reionization in particular.

Numerous theoretical models for cosmic reionization have been
constructed, often based on semi-analytic models of the reionization
process or simple radiative transfer postprocessing of dark matter
simulation outputs. \citet{Furlanetto2004} developed an excursion set
approach to reionization that has seen widespread use in analytic and
semi-numerical models of reionization \citep[e.g.][]{Alvarez2007,
  Zahn2011, Mesinger2011, Raicevic2011, Battaglia2013}. 
Many different numerical algorithms for direct
radiative transfer simulations have also been developed over the
years, in most cases however based on static density fields derived
from dark matter only simulations or from simplified hydrodynamic
simulations \citep[e.g.][]{Sokasian2001, Ciardi2003, Iliev2006,
  McQuinn2007, Zahn2007, Croft2008, Trac2008, Aubert2010, Ahn2012}.

Recently, full radiation hydrodynamics simulations that follow cosmic
reionization and structure formation simultaneously and
self-consistently have become possible. These calculations can in
principle account for radiative feedback processes on forming
galaxies, for example from inhomogeneous photoionization
heating. Pioneering work of this type has been presented by
\citet{Gnedin2000}, but only in recent years it has become possible to
study approximately representative cosmological volumes in this way
\citep[e.g.][]{Petkova2011b, Pardekooper2013}.

Some of the most advanced studies of this kind include the simulations
recently presented by \citet{Norman2013} and \citet{So2014}, who use
full radiation hydrodynamics simulations of cosmic structure formation
on a uniform grid. A fixed spatial grid resolution does however not
allow a proper resolution of internal galaxy structure, which
compromises the ability of the simulations to reliably predict the
build up of stellar mass. To remedy this problem, \citet{Gnedin2014a}
and \citet{Gnedin2014b} employ adaptive mesh refinement techniques and
simulate galaxy formation in cosmological volumes with high spatial
and mass resolution. Similar work has recently been
presented by \citet{Pawlik2015}, based on hydrodynamical SPH
simulations coupled self-consistently to the radiative transfer scheme
{\small TRAPHIC} \citep{Pawlik2008}. However neither group evolves the
simulations significantly past the EoR (let alone to $z=0$)
due to the high computational cost involved, so it is not yet clear
whether these simulation models would also yield a plausible {\em present-day}
galaxy population.

This body of theoretical works has made it clear that the source
population primarily responsible for reionization is most likely
star-forming small galaxies at high redshift. While so-called Pop-III
stars may boost the high redshift photon production rate, the overall
contribution of these ``first star'' sources is likely only of secondary
importance \citep{Pardekooper2013, Wise2014}. Similarly, another potential
source of ionizing photons, active galactic nuclei (AGN), is not expected
to be critical at high redshift due to their large mean separation
\citep{Hopkins2007, Faucher-Giguere2009} and still fairly limited
cumulative luminosity. Instead, it is often argued that small
proto-galaxies, with stellar masses just around $10^4\,{\rm M}_\odot$ or
even lower may dominate the ionizing budget \citep{Pardekooper2013}.
\citet{Ahn2012} showed that these mini halos alone can not
complete reionization, but significantly contribute towards an earlier
onset of reionization and thus enhance the optical depth towards the last
scattering surface. These small halo mass systems are further assisted by
suggestions that the escape fraction may strongly rise towards small halo
masses, as inferred by \citet{Wise2014} based on radiation hydrodynamics
simulation of faint high redshift galaxies. Using this finding in a
semi-analytic model for reionization, \citet{Wise2014} have also
demonstrated that the first galaxies may plausibly constitute the
reionizing sources, yielding an optical depth consistent with
\citet{Planck2013CosmoParams} without exceeding the UV emissivity
constraints by the Ly-$\alpha$ forest.

It is interesting however to note that the cosmic star formation history
inferred from observations is predicted to rapidly decline towards high
redshift \citep[e.g.][]{Ellis2013}. This is also expected
\citep{Hernquist2003} and desirable on theoretical grounds
\citep{Scannapieco2012}, because simulation models of galaxy formation
need to resort to an extremely efficient suppression of high-redshift star
formation in order to successfully describe present-day galaxy properties
\citep[e.g.][]{Stinson2013}. A very low level of high redshift star
formation is however quite the opposite of what seems necessary to explain
early reionization and the comparatively high optical depth inferred from
CMB experiments. This tension makes it difficult to attribute reionization
entirely to young galaxies with more or less ordinary stellar populations.
In $\Lambda$CDM models with low normalization $\sigma_8$, or in
alternative warm dark matter cosmologies, this problem is further
exacerbated \citep{Yoshida2003a, Yoshida2003b}, whereas in certain
non-standard dark energy models that shift structure growth to earlier
times, for example `early dark energy' \citep{Wetterich2004, Grossi2009},
it may be alleviated.

Recently, cosmological hydrodynamic simulations of galaxy formation such
as the {\em Illustris} \citep{Vogelsberger2014} or {\em Eagle}
\citep{Schaye2014} projects have advanced to a state where they produce
realistic galaxy populations simultaneously at $z=0$ and at high
redshifts, throughout a representative cosmological volume. This is
achieved with coarse sub-resolution treatments of energetic feedback
processes that regulate star formation, preventing it from exceeding the
required low overall efficiency. Ideally the process of reionization
should be coupled dynamically to such a galaxy formation model. However,
running several such reionization simulations for different escape
fraction parameterizations at the resolution and size of Illustris would
be computationally very expensive. Thus we have here reverted to study
reionization in post processing only. This approach is not as
self-consistent as the most recent generation of full
radiation-hydrodynamics simulations of galaxy formation
\citep[][]{Gnedin2014a, Norman2013, Pawlik2015}, but it can take advantage
of a successful description of the evolving source population and the
intervening gas distribution down to low redshift.

An important manifestation of feedback are galactic winds and outflows
that substantially modify the distribution of the diffuse gas in the
circumgalactic medium (CGM) and the IGM. This in
turn also influences the gas clumping and the recombination in models of
cosmic reionization. It thus becomes particularly interesting to test
whether detailed models of galaxy growth such as Illustris are in
principle also capable of delivering a successful description of cosmic
reionization, and if so, what assumptions are required to achieve such a
success.

This is exactly the goal of this paper. We use a sequence of snapshots
with high time resolution of the high-resolution Illustris simulation and
combine them with a radiative transfer scheme that is capable of
accurately evolving ionizing radiation for an arbitrary number of sources.
We are particularly interested in the question of whether the star
formation history predicted by Illustris can reionize the universe early
enough to be consistent with observational constraints, and how the
reionization transition proceeds in detail in this scenario. Because we
have implemented two different radiative transfer methods, we can also
evaluate how well they intercompare, thereby providing an estimate for
systematic uncertainties related to these radiative transfer methods. We
also explicitly test the impact and accuracy of the often adopted reduced
speed of light approximation.

This work is structured as follows. In Section~\ref{sec:methods}, we
introduce the Illustris simulation and our different radiative
transfer schemes that we use to follow cosmic reionization. We then
turn to a discussion of our primary results in
Section~\ref{sec:results}. In Section~\ref{sec:numerics}, we analyse
numerical caveats such as resolution dependence or systematic effects
due to different radiative transfer approximations and discuss our
findings. This is followed by our conclusions in
Section~\ref{sec:conclusion}.

\section{Methods}  \label{sec:methods}

\subsection{The Illustris simulation}

Recently, \citet{Vogelsberger2014} introduced the Illustris simulation
suite, an ambitious attempt to follow cosmological hydrodynamics and the
feedback processes associated with galaxy formation in a sizable region of
the universe. The highest resolution simulation of the project employed
$2\times 1820^3$ particles and cells in a box 106.5 Mpc across, yielding a
mass resolution of $1.26\times 10^6\,{\rm M}_\odot$ in the baryons, and
$6.26\times 10^6\,{\rm M}_\odot$ in the dark matter. The cosmology adopted
is given by $\Omega_m =0.2726$, $\Omega_0 =0.7274$, $\Omega_b =0.0456$,
$\sigma_8 = 0.809$, $n_s = 0.963$, and $H_0 = 70.4\,{\rm
km\,s^{-1}Mpc^{-1}}$, which is consistent with the most recent
determinations from WMAP9 and Planck. The simulations employed the
moving-mesh code {\small AREPO} \citep{Springel2010}, which is well-suited
to applications in cosmic structure formation.

The physics model employed by Illustris includes radiative cooling,
metal enrichment based on 9 elements, star formation, stellar
evolution and mass return, supernova feedback by means of a kinetic
wind feedback, and black hole growth and associated feedback processes
in a quasar- and radio-mode.  We refer to \cite{Vogelsberger2013} and
\cite{Torrey2014} for a description of the full details and basic
tests of the model. A number of different studies have analysed
structure formation in Illustris, making it clear that many basic
properties of the observed galaxy populations are approximately
reproduced by the simulation model. This in particular includes
constraints on the stellar mass function at different epochs
\citep[]{Vogelsberger2014b, Genel2014}, the morphologies and spectra
of galaxies \citep{Torrey2015}, the colors of satellite systems
\citep{Sales2015}, the stellar halos of galaxies
\citep{Pillepich2014}, the nature of high redshift, compact galaxies
\citep{Wellons2014}, the galaxy-galaxy merger rate
\citep{Rodriguez-Gomez2015}, the kinematics and metal abundance of
damped Lyman-alpha absorbers \citep{Bird2014}, or the evolution of the
black hole mass density and quasar luminosity function
\citep{Sijacki2014}. The galaxy formation predictions by Illustris are
hence in broad agreement with observations, which adds additional
motivation to ask whether they at the same time yield a plausible
reionization history.

A self-consistent treatment of the UV background using radiative transfer
was however not included in Illustris, as this is still beyond reach in
such large cosmological simulations that are evolved to low redshift.
Instead, an external, spatially uniform and time-dependent UV background
was imposed based on the model of \citet{Faucher-Giguere2009}, and dense
gas is self-shielded from UV background radiation using a prescription
derived from \citet{Rahmati2013}. A coarse treatment of an AGN proximity
effect was included where accreting AGNs modify the ionization balance of
gas in their environment \citep{Vogelsberger2013}. In this work, we
therefore aim to study reionization through postprocessing of the
Illustris simulation, making use of the significant number of output dumps
\citep[more than 128, with an output spacing as in the Aquarius
project,][]{Springel2008} stored for the calculation. This allows us to
take the temporal information about the growth of cosmic structures into
account, avoiding the simplification of a static density field often
adopted in past work. Another advantage of Illustris is the reasonably
large box size of 106.5 Mpc. While a still larger volume would clearly be
desirable, studies of cosmic variance of reionization have suggested that
$\sim 100\,h^{-1}\mathrm{Mpc}$ corresponds to the minimum box size
required to obtain a reliable mean reionization history
\citep{Mesinger2007, Iliev2014}. The volume available in Illustris falls
slightly short here, but is approximately still sufficient for hydrogen
reionization. Studying HeII reionization with Illustris would be more
problematic however due to the incomplete sampling of bright and rare
quasars.

\subsection{Radiative transfer}

The governing equation of radiation transfer is the Boltzmann
equation for the photon distribution function, 
\begin{align}
 \frac{\partial f_{\gamma}}{\partial t} + \frac{\partial}{\partial \vect{x}}  \left (\frac{c}{a} \vect{n} f_{\gamma}\right )  = \left .\frac{\partial f_{\gamma}}{\partial t} \right |_{\rm sources}  - \left . \frac{\partial f_{\gamma}}{\partial t} \right |_{\rm sinks}, 
\label{eqn:rad}
\end{align}
with cosmological scale factor $a$ and speed of light $c$. The function
$f_{\gamma} (\vect{x}, \vect{n}, \nu,t)$ describes the number density of
photons of frequency $\nu$ at time $t$, propagating in the direction of
the normal vector $\vect{n}$. In the following, we will replace the
frequency dependence by one effective frequency bin that accounts for all
hydrogen ionizing photons, for simplicity. This can in principle be
straightforwardly generalized to multiple frequency bins, as is necessary,
for example, to also treat helium reionization. In
equation~(\ref{eqn:rad}), we ignore the redshifting of ionizing photons
which is justified if photons are destroyed in ionization events shortly
after their creation.

Solving equation~(\ref{eqn:rad}) in full generality is very difficult,
hence a large variety of approximation methods for cosmological
radiative transfer have been developed, each with different advantages
and shortcomings that ultimately dictate the regimes where they are
applicable. Many radiative transfer schemes are based on the idea of
characteristics \citep{Mihalas1984}, where the optical depth is
individually integrated along rays between different computational
cells. As this operation quickly becomes extremely costly many
different techniques have been developed to accelerate the
calculations, for example by using short-characteristics
\citep[e.g.][]{Nakamoto2001}, or by hierarchically splitting rays
\citep{Abel2002, Trac2007}. Ray-tracing schemes are particularly well
suited for dealing with a small number of isolated sources, and in that
sense are ideal for following the growth of an isolated ionized region
around a point source \citep{Abel1999}. However, the reionization
problem involves a large number of sources (all the stars in a large
number of galaxies), favouring the use of other methods.

One popular approach is to take moments of the radiative transfer
equation, leading to an evolution equation for the mean specific intensity
that needs to be closed with an estimate of the local Eddington tensor
\citep{Gnedin2001, Aubert2008, Petkova2009, Finlator2009}. The simplest
variant of this approach is flux-limited diffusion (FLD) of radiation
\citep{Levermore1981, Turner2001}, which works particularly well in the
optically thick regime but may fail badly in situations where the optical
depth is low and shadowing might be important. Better accuracy can be
obtained in moment-based approaches when the local Eddington tensor can be
estimated with reasonable accuracy. This is attempted, for example, in the
optically thin variable Eddington tensor (OTVET) approximation of
\citet{Gnedin2001}, or by invoking other simple approximations such as the
so-called M1-closure \citep{Levermore1984, Aubert2008}.

More general radiative transfer such as Monte Carlo methods
\citep[e.g.][]{Maselli2003, Nayakshin2009} may also be invoked, but
their computational cost is comparatively large making it difficult to
avoid a high level of noise in large-scale simulated radiation fields.
The {\small TRAPHIC} approach of \citet{Pawlik2008} improves on this
by restricting the transport of photon packets to a finite set of
angular cones, while retaining the flexibility of the Monte Carlo
scheme. A conceptually similar idea is followed in the radiation
advection scheme of \citet{Petkova2011}, where the radiation field is
directly discretized in angular space and transported with a
conservative advection solver on an unstructured grid.
 
In our work here, we consider two different numerical methods to solve
the advection part of the radiation transport, described by the left
hand side of equation~(\ref{eqn:rad}). The first method is the
cone-based advection method proposed by \citet{Petkova2011}, and the 
second method is a simpler moment-based approach where
the M1 closure for the Eddington tensor is used \citep{Aubert2008,
  Rosdahl2013}. This allows us to assess how strongly fundamental
differences in the numerical radiative transfer approximation affect
the predicted reionization transition, thereby yielding an estimate
for this contribution to the systematic uncertainties in theoretical
EoR predictions.

\subsection{Cone-based advection method}

In the cone based method introduced by \citet{Petkova2011}, the angular
dependence of the distribution function $f_{\gamma}$ is discretized in
cones of equal opening angle. We use the {\small HEALPIX} tessellation
scheme \citep{Gorski2005} to subdivide the unit sphere, giving
$N_{\mathrm{pix}} = 12\times 4^n$ cones of equal solid angle, where $n$ is
an integer parameter determining the resolution level. The spatial
dependence can be discretized using either a structured or unstructured
mesh. As we are mostly interested in volume-weighted effects of
reionization (such as the volume filling factor of ionized gas) and want
to efficiently exploit GPUs as computational accelerators we adopt a
Cartesian grid in this work. This leaves us with $N_{\mathrm{pix}}$ photon
fields, $f_l \left(\vect{x}_{i,j,k}\right)$, at grid coordinates
$\vect{x}_{i,j,k}$, with $f_{\gamma} = \sum f_l$ being the total
angle-integrated intensity.

Leaving the source terms aside for the moment (which are treated in an
operator-split fashion), the cone-based method solves a hyperbolic advection
equation,  
\begin{align}
    \frac{\partial f_l}{\partial t} + \frac{\partial}{\partial \vect{x}}  \left (\frac{c}{a} \vect{n}_l f_l \right )  = 0,
\end{align}
for each of the angular-decomposed photon fields $f_l$.  The principal
transport direction $\vect{n}_l$ of cone $l$ is set by the {\small
  HEALPIX} tessellation of the unit sphere. However, to make sure that
the entire cone of each tile is illuminated homogeneously and that the
transport does not excessively collimate the radiation into an angle
smaller than the prescribed angular resolution, $\vect{n}_l$ is obtained by taking
\begin{align}
 \vect{n}' = - \frac{\nabla f_{\gamma}}{\left | \nabla f_{\gamma} \right | } ,
\end{align}
and additionally, if this direction should fall outside of a cone with
center $\vect{\tilde{n}}_l$ and opening angle
$\phi_{\mathrm{max}} = \sqrt{4/N_{\mathrm{pix}}}$, it is projected
back inside that cone. Here $\vect{\tilde{n}}_l$ is the geometric
center of the $l^{\rm th}$ {\small HEALPIX} cell. In other words, the
advection direction is constrained to lie within the cone but
otherwise tries to reduce gradients in the photon intensity across the
cone, thereby ensuring a homogenous illumination of the cone.

We typically integrate the above transport equations using a reconstructed
piecewise constant photon intensity field combined with an upwind scheme
to determine the photon flux between cells. We use explicit time
integration and hence need to restrict the timestep to something of order
$h/c$, where $h$ is the cell radius and $c$ the speed of light. This leads
to quite small timesteps. This could be mitigated by invoking a reduced
speed of light approximation \citep{Gnedin2001}, but as we discuss in
section~\ref{sec:rsl} we have not used this here to avoid any inaccuracies
that can be introduced by this approximation.

\subsection{Moment method with M1 closure}

For comparison with the angular resolved transport scheme, we also
implemented a moment based advection scheme. Here the radiative
transfer equation is discretized by taking the first two moments of
equation~(\ref{eqn:rad}):
\begin{align}
  \frac{\partial N}{\partial t} &+  \nabla \vect{F} = 0,\\
  \frac{\partial \vect{F}}{\partial t} &+ c^2 \nabla \mathbf{P} = 0,
\end{align}
where the photon density $N$, the flux vector $\vect{F}$, and the
photon pressure tensor ${\mathbf P}$ have been introduced. Solving
this system of equations requires a closure for the radiation pressure
tensor $\mathbf P$. The simplest closure is to assume an isotropic
pressure, which yields the family of flux-limited diffusion methods
\citep{Levermore1981}. Another possibility is the OTVET approximation
where a preferred local streaming direction of photons is estimated
based on the source locations and their strengths
\citep{Gnedin2001}. More accurate estimates of a variable Eddington
tensor (VET) may for example be obtained by employing a special
short-characteristics method to estimate the local Eddington tensor
\citep{Davis2012, Jiang2014}.

An alternative is the M1 method, where a fully local closure is used
instead \citep{Levermore1984, Aubert2008}. This greatly simplifies the
computations because information about the surrounding regions does
not explicitly have to be taken into account. The photon pressure
tensor of the M1 closure can be parametrized as:
\begin{align}
  \mathbf{P} = \left ( \frac{1 - \xi}{2} \mathbf{I} + \frac{3 \xi -1}{2} \vect{n} \otimes \vect{n} \right ) N,
\end{align}
with 
\begin{align}
\vect{n} = \frac{\vect{F}}{\left|\vect{F}\right|}, \quad
\xi = \frac{3+4f^2}{5+2\sqrt{4-3f^2}}, \quad f = \frac{\left|\vect{F}\right|}{cN},
\end{align}
and identity matrix $\mathbf{I}$. Here $f$ essentially determines how
directed the flux is, interpolating between the two limiting cases of
radiation diffusion and photon streaming. The parameter $\xi$ varies
smoothly from $\frac{1}{3}$ to $1$ between these cases. In the case of
$\xi=\frac{1}{3}$, which occurs for an undirected flux, the result is
isotropic advection. The other limit of $\xi=1$ corresponds to fully
directed transport with one dominant propagation direction.

While in practice the M1 closure often produces surprisingly accurate
results, it is easy to construct situations where it fails. For
example, one obvious shortcoming occurs when two light beams directly
encounter each other from opposing directions. As the photon field is
essentially described by M1 as a collisional fluid with a restricted
form for the local Eddington tensor, a `scattering' at the beam
intersection point is unavoidable, resulting in an unphysical
propagation direction. Nevertheless, one can hope that in many
practical applications such inaccuracies occur sufficiently rarely
that the overall results are still reliable. Whether or not this is
really the case is problem dependent and needs ultimately be tested by
comparing with more accurate techniques that are based on different
approximations, something that we also pursue in this work.

\subsection{Ionization network and time integration}

The evolution of the ionization state of hydrogen is given by the
following system of equations:
\begin{align}
\frac{{\rm d} n_{\gamma}}{{\rm d} t} &= - c \sigma n_{\rm H} x n_{\gamma} + \left[\alpha_{\mathrm{A}}(T) - \alpha_{\mathrm{B}}(T) \right] n^2_{\rm H} \left (1-x \right)^2,\\
\label{eqn:ion}
\frac{{\rm d} x n_{\mathrm H}}{{\rm d} t} &= \alpha_{\mathrm{A}}(T) n^2_{\rm H} \left (1-x \right)^2- \beta(T)  n^2_{\rm H} x (1-x) - c \sigma n_{\rm H} x n_{\gamma},
\end{align}
with the hydrogen number density $n_{\mathrm H}$ , the frequency averaged
ionization cross section $\sigma$, the number density of ionizing photons
$n_{\gamma}$ and the neutral hydrogen fraction $x$. The case A and case B
recombination rates $\alpha_{\mathrm{A}}$ and $\alpha_{\mathrm{B}}$ and
collisional ionization rate $\beta$ are taken from \cite{Hui1997}. The
chemical network is complemented by source terms in the thermal energy
evolution describing cooling and photoionization heating:
\begin{align}
 \frac{{\rm d} u}{{\rm d} t} = \epsilon_{\gamma} c \sigma n_{\rm H} x n_{\gamma}     -{\cal C}(T),   \label{eqn:temp}
\end{align}
where $\epsilon_{\gamma} $ is the average gain in thermal energy per
ionization event. The cooling rate ${\cal C}(T)$ includes cooling due
to recombination, collisional ionization, collisional excitation and
Compton cooling by scattering on CMB photons.

We solve the chemical network following the method outlined in
\citet{Rosdahl2013}. First, the photon number $n_{\gamma}$ is updated by
an implicit Euler step, then the temperature $u$ and finally the
ionization state $x$. Each partial update step is implicit in the already
updated quantities and the currently updated one. The full radiative
transfer equation~(\ref{eqn:rad}) is then solved by Strang-like operator
splitting, computing first the left hand side for half a time step
(advection), then applying the source and sink terms, and finally
completing the timestep by another half advection step. The two advection
half steps themselves are solved by dimensional splitting.

In our cone-based advection method, we usually use the $n=1$ {\small
HEALPIX} tessellation level with $N_{\mathrm{pix}} = 48$ pixels. Due to
the large number of photon fields the computations are demanding in memory
as well as in computational power. To speed up our simulations, we are
employing GPUs, an approach similar in spirit to \citet{Aubert2010}. We
use a uniform Cartesian mesh which makes our algorithm ideal for GPU
computing, as this leads to a regular memory access pattern and
facilitates the use of groups of threads with the same execution path.
Details of our GPU implementation will be described elsewhere (Bauer et
al., in preparation).

\subsection{Reionization in post processing}
In this work, we study the progress of cosmic reionization by following
the time-resolved evolution of the density field and the source population
of the Illustris simulation in postprocessing. The unstructured Voronoi
mesh that stores the baryon distribution in the simulation outputs needs
to be rebinned onto a regular Cartesian mesh to allow use of our GPU-based
code. To this end, we assign the mass of each Voronoi cell onto the
density grid using a spline-kernel assignment. This yields a less noisy
and smoother density field than obtained, e.g., by using a clouds-in-cells
(CIC) or nearest-grid-point (NGP) assignment kernel \citep{Hockney1981}.
The actual density field used in our reionization calculation is then
continuously updated by linearly interpolating in time between the two
nearest binned density grids available in $\log a$ space, where $a=
1/(1+z)$ is the cosmological scale factor. The Illustris outputs are
spaced roughly 65 Myrs apart at the relevant redshift $z\sim 6$, allowing
us to bin the density field in total 35 times between redshift $z=21.8$
and $z=4.9$. Using a set of small subboxes cut out from Illustris and
stored with much higher time-resolution we have checked that the sparser
time resolution of the main outputs is still reasonably accurate. It
shifts the completion of reionization towards slightly earlier times, but
the uncertainty in our results is still dominated by the parametrization
of the escape fraction. Initially we start with a uniform temperature
field with $T = 100\, \mathrm{K}$ and then follow the temperature
evolution using equation~(\ref{eqn:temp}), ignoring the intrinsic
temperature evolution of the underlying Illustris simulation. We note that
in this paper we consider hydrogen reionization only.

\subsection{Escape fraction}

\begin{table*}
\begin{tabular}{lccccc}
\hline
\multicolumn{4}{l}{Overview of radiative transfer simulation models}\\ 
\hline
Name & Resolution & $f_{0}$ & $\kappa$ & Advection method  \\
\hline

V1\_{\em X}\_CONE & $256^3$ \dots $512^3$ & $0.04$ & $4$ & Cone based method\\
V5\_{\em X}\_CONE & $256^3$ \dots $512^3$ & $0.04$ & $3.6$ & Cone based method\\
C2\_{\em X}\_CONE & $256^3$ \dots $512^3$ & $0.2$ & $0$ & Cone based method\\

V1\_{\em X}\_M1 & $256^3$ \dots $1024^3$ & $0.04$ & $4$ & M1 method\\
V5\_{\em X}\_M1 & $256^3$ \dots $1024^3$ & $0.04$ & $3.6$ & M1 method\\
C2\_{\em X}\_M1 & $256^3$ \dots $1024^3$ & $0.2$ & $0$ & M1 method\\
\hline
\end{tabular}
\caption{Overview of the different radiation transfer calculations performed for this study. We carried out runs with a resolution ranging from $256^3$ up to $1024^3$ cells, which 
  is indicated by replacing the placeholder `{\em X}' with the number of cells per dimension in the actual run name.
  We compare different escape fraction parameterizations, characterized by $f_0$ and $\kappa$. The three different choices we adopted for this are labeled  `V1', `V5' and `C2' in the simulation names. Finally, 
  for each of the models we compare two different advection schemes for the radiation, one based on the M1 closure relation, the other on an explicit discretization of the solid angle (`cone based') . \label{tab:sim}}
\end{table*}

Of all ionizing photons emitted by stars, only a fraction
$f_{\mathrm{esc}}$ reaches the IGM, while the other photons are absorbed
by the denser interstellar medium (ISM) or by dust. In principal, the
escape fraction depends on individual halo properties like halo mass or
dust content. Unfortunately, only little is known about the real values of
the escape fractions, especially at high redshifts, making this parameter
one of the primary uncertainties in studying the EoR. However, theoretical
models have started to constrain the escape fraction, albeit with large
systematic uncertainties. For example, the simulations of \citet{Wise2014}
suggest a sharp increase of the escape fraction towards smaller masses,
reaching $50\%$ at halo masses of $10^7\,{\rm M}_\odot$.

The simplest model is evidently to adopt a globally constant escape
fraction that is the same for all galaxies at a given epoch. This is what
we shall assume here, because one may well argue that in light of the many
other uncertainties adopting a more complicated model would not be
justified \citep[see also the discussion in][]{Pawlik2015}. As a default
choice for the constant escape fraction model we have considered
$f_{\mathrm{esc}} = 0.2$ (C2 model). This almost certainly constitutes an
overestimate for low redshift galaxies, but is perhaps not overly
optimistic at high redshift.

Besides such a globally constant escape fraction, we also consider an
evolution of the $f_{\mathrm{esc}}$ value with time. To this end, we
use the model of \citet{Kuhlen2012} who proposed an evolution of the
escape fraction as a function of redshift according to: \begin{align}
  f_{\mathrm{esc}} = \min\left[1,f_0 \left ( \frac{1+z}{5} \right
    )^{\kappa} \right]. \end{align} Our default choices for the
parameters $f_0$ and $\kappa$ are $f_0 = 0.04$ and $\kappa = 4$,
implying a rise of the escape fraction with redshift (V1 model). We
have also calculated results for a variety of other fiducial parameter
choices as well. For an overview see Table~\ref{tab:sim}.

We stress again that the escape fraction is essentially a free
parameter in our treatment. Its interpretation is complicated by the
fact that the absorbing ISM is not totally absent in our reionization
simulations. It is just severely underesolved, and thus some part of
the boost in recombination rates due to a highly clumped environment
is missing.  However, photons are still consumed for (repeatedly)
ionizing the material in these high density regions. Unresolved small galaxies and thus missing UV photons might be compensated by a slightly larger escape fraction than otherwise would be needed.

\subsection{Source modelling}

\begin{figure}
\begin{center}
\vspace*{-0.5cm}\hspace*{-0.5cm}%
\includegraphics{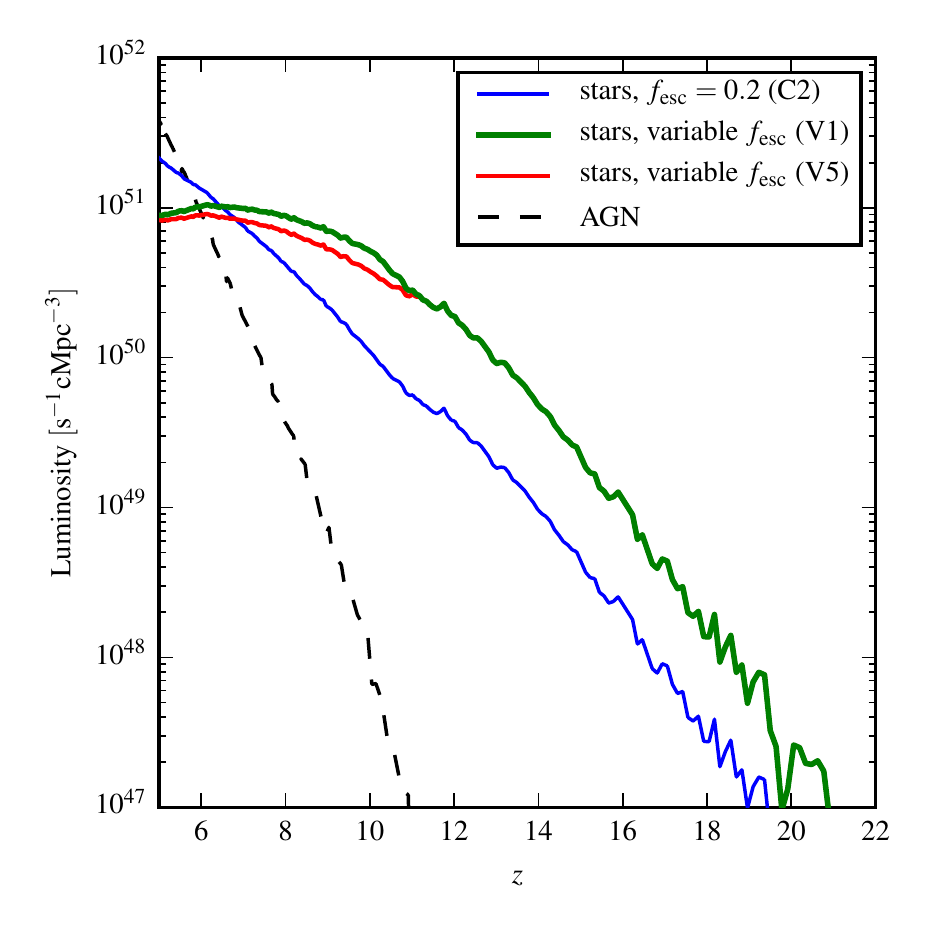}%
\vspace{-0.5cm}%
\caption{Time evolution of the total ionizing luminosity resulting for
  our different escape fraction models. The model with a constant
  escape fraction (blue lines) tracks the shape of the cosmic star
  formation history. Our default models with a redshift dependent
  escape fraction are given by the green and red lines. For comparison, we
  also include the ionizing luminosity implied by the quasars included
  in our simulation (dashed black line), adopting a simple conversion of quasar
  accretion rate to ionization radiation output. This demonstrates
  that the quasar contribution picks up too late to cause the initial
  hydrogen reionization, but it may play a role in keeping the
  universe ionized at later times, as well as for completing late-time
  helium reionization. \label{fig:lum}}
\end{center}
\end{figure}

The main source of ionizing photons considered in our model is ordinary
stellar populations in young stars, which arguably appear to be the most
likely source responsible for reionization. In this work we are mainly
interested in testing this hypothesis based on directly adopting the
stellar populations forming in the Illustris simulation. Their ionizing
luminosities as a function of stellar age and metallicity are taken from
{\small STARBURST-99} \citep{Leitherer1999}. Figure~\ref{fig:slices} gives
a visual impression of the binned gas density field in Illustris at $z=7$,
and of the clustered galaxy population that represents our source
population. Because the local ionization rate can change on much shorter
timescales than the density, we bin the luminosity field with much finer
time resolution than the density field. In our default set-up, we binned
it 200 times during the duration of the reionization simulation including
the actual birth time of the stellar sources. The actual luminosity used
in the simulation to integrate the source terms is then again interpolated
in $\log a$ space from this large grid of luminosity density fields. The
ionizing luminosity of a single stellar source is always distributed in a
photon conserving way to the radiation grids. The assignment of the
luminosity is done using a CIC interpolation scheme.

The ionizing sources we have at our disposal from the main Illustris
simulation are related to the star formation rate and black hole accretion
rate densities. The corresponding time evolution of the net ionizing
luminosity density is shown in Figure~\ref{fig:lum}, together with
different scenarios for the escaping luminosity according to our escape
fraction models. The blue line shows our constant escape fraction model
(C2 model), while the green line is our default model for a time-varying
escape fraction (V1 model). The red line (V5 model) represents a variable
escape fraction model with a different value for the exponent $\kappa$
than in the V1 model. These models rapidly rise for some time, but at
around a redshift of $z \simeq 8$, the escaping radiation actually reaches
a maximum and then starts to slowly decline again. Here the strong
decrease of the escape fraction from $1$ to $0.04$ marginally
over-compensates the further increase of the star formation density and
the raw ionizing luminosity. We note that the maximum coincides with the
epoch where we expect most of the hydrogen to be reionized.

An alternative source of ionizing radiation is in principle provided
by AGNs. We assume a bolometric AGN luminosity described by
\begin{align}
  L_{\mathrm{bol}}^{\mathrm{AGN}} = \left(1 - \epsilon_{\mathrm{f}}\right) \tilde{\epsilon}_{\mathrm{r}} \dot{M}_{\mathrm{BH}} c^2,
\end{align}
with a radiative efficiency of $\epsilon_\mathrm{r} = 0.2$ and an
energy fraction of $\left(1-\epsilon_{\mathrm{f}}\right) = 0.95$
available for radiation. The luminosity is converted into a rate of
ionizing photons assuming a parameterized AGN SED \citep{Korista1997}
equal to
\begin{align}
  f^{\mathrm{AGN}}(\nu) = \nu^{\alpha_{\mathrm{UV}}} \exp\left(-\frac{h \nu}{k T_{\mathrm{BB}}}\right) \exp \left( - \frac{10^{-2} \mathrm{Ryd}}{h \nu} \right) + a \nu^{\alpha_{\mathrm{x}}},
\end{align}
with a suppression of the UV component at a temperature of
$T_{\mathrm{BB}} = 10^6\,\mathrm{K}$, a UV component slope of
$\alpha_{\mathrm{UV}} = -0.5$ and an X-ray component slope of
$\alpha_{\mathrm{X}} = -1$. To obtain an approximation for the maximum
contribution to reionization, we assume an escape fraction of
$f_{\mathrm{esc,AGN}} = 1$. For comparison, we show the AGN ionizing
luminosity as a dashed black line in Figure~\ref{fig:lum}. The AGN
contribution only becomes relevant at around $z=6$, but by then most of
the hydrogen must already be ionized, making it unlikely that AGNs are
significantly contributing to the initial EoR transition. However, they
might play a role in keeping the universe ionized at a later time, and
almost certainly are important for completing helium reionization at lower
redshift, thanks also to their harder spectrum. As we only study hydrogen
reionization, in the following only stellar emission is considered as
sources of ionizing photons.

\section{Simulation results}  \label{sec:results}

An overview of the various reionization simulations carried out in
this work is given in Table~\ref{tab:sim}. To examine resolution
dependences, we computed several of our models with grid resolutions of
up to $1024^3$ cells. We considered two different models with a
time-variable escape fraction and contrasted them with one model with
a fixed escape fraction. In order to asses the impact of different
radiative transfer methods we performed most of our simulations both
with our default cone-based approach as well as with the moment-based
method with M1 closure.

\begin{figure*}
\begin{center}
\setlength{\unitlength}{1cm}
\vspace*{-2cm}\includegraphics{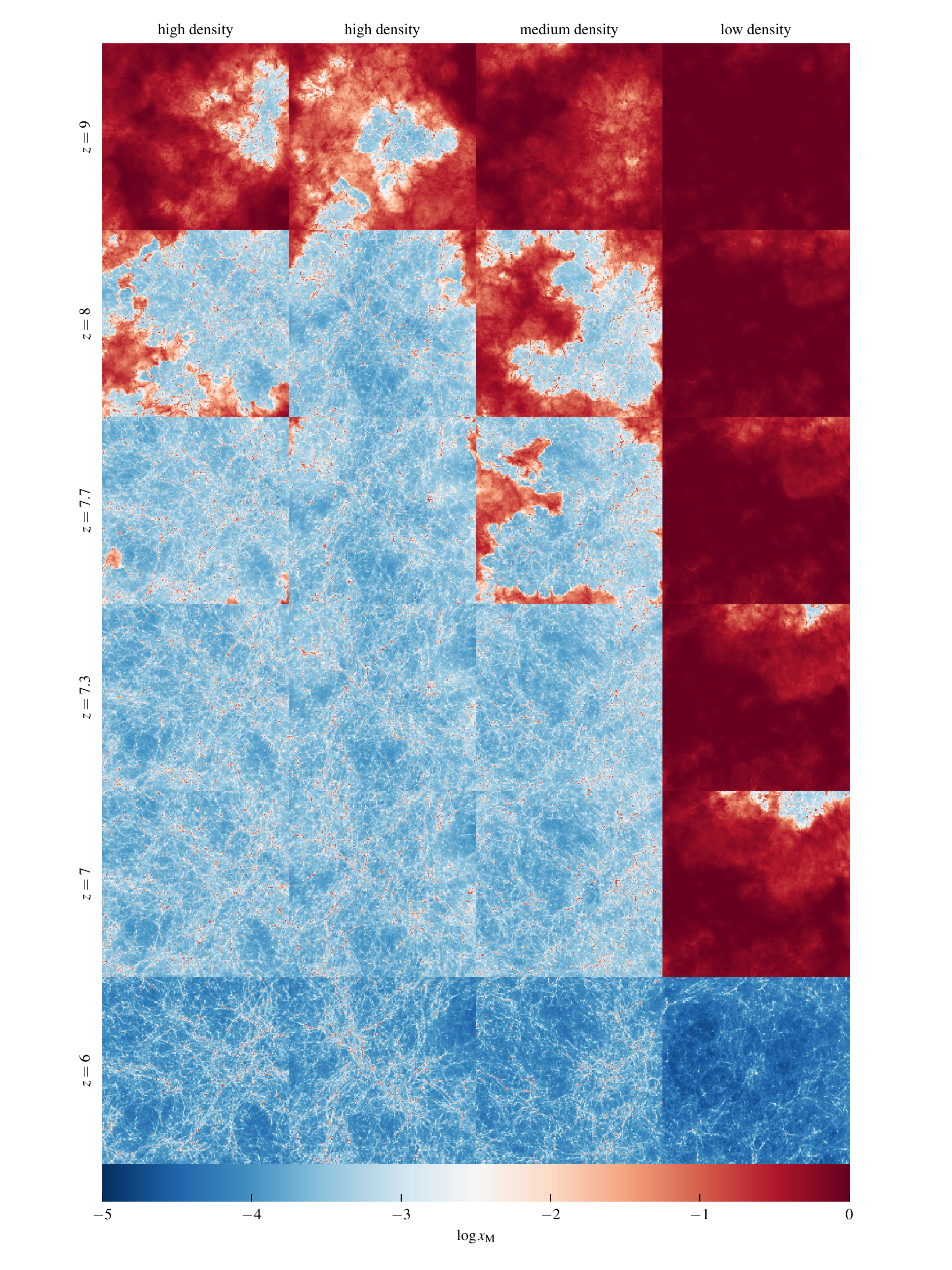}
\caption{Progression of reionization as seen in the neutral hydrogen
  fraction in slices through selected
  sub-volumes in Illustris, each with a side-length of $21.3\,{\rm cMpc}$. 
  Each column shows the time evolution of a different, 
  randomly selected environment in our V1\_1024\_M1 model; 
  the two columns on the left
correspond to an average density higher than the mean, the other
columns have medium and
  low mean density, as labeled. Each row gives a
  different redshift, from $z=9$ (top) to $z=6$ (bottom). 
  \label{fig:IonizationSlices}}
\end{center}
\end{figure*}

The progress of reionization in different environments is visually
shown in Figure~\ref{fig:IonizationSlices}, where we compare two
regions around very massive halos with a more average environment
around a typical medium-sized halo, and an underdense region. The
different projections show the four regions at six different output
times. Reionization starts inside the most massive halos first, and
then quickly ionizes the surrounding regions. Compared to such a
high-density environment, the onset of reionization is considerably
delayed around a medium mass halo. More drastically, the underdense
region only begins to be reionized once the denser regions have almost
completed their reionization transition. The visual impression is thus
qualitatively consistent with an inside-out reionization scenario in
which halos in high-density regions are affected first and lower
density voids are reionized rather late, for the most part after
overdense gas been reionized \citep{Razoumov2002}. This contrasts with
suggestions that low density regions are reionized quite early and
only then the reionization fronts progress to ionize filaments and gas
in halos \citep{Gnedin2000}.

\subsection{Reionization history}

\begin{figure}
\begin{center}
\setlength{\unitlength}{1cm}
\hspace*{-0.8cm}
\resizebox{9.6cm}{!}{\includegraphics{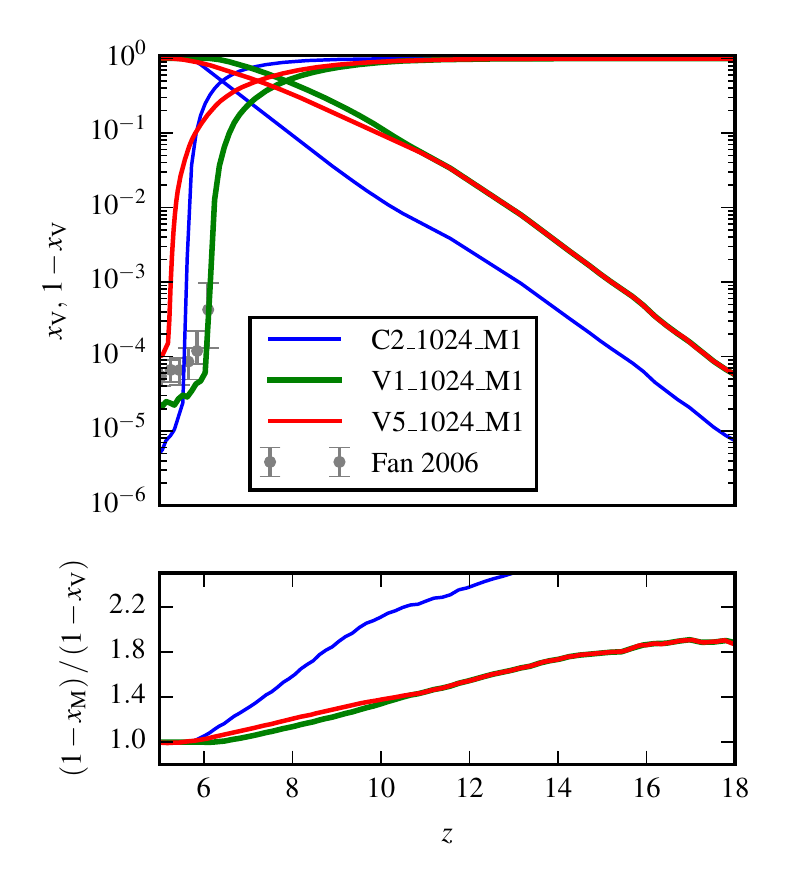}}%
\vspace*{-0.7cm}
\end{center}
\caption{The ionization history for different escape fraction
  parameterizations. The upper panel shows the evolution of the volume
  weighted ionization fraction (solid lines rising to low redshift) and the
  neutral hydrogen fraction (lines dropping towards low redshift), as
  a function of redshift. For comparison, observational constraints by
  \citet{Fan2006a} are shown as symbols with error bars. The lower
  panel shows the ratio between mass- and volume-weighted ionization
  fractions $\left(1-x_{\mathrm{M}}\right)/\left(1-x_{\mathrm{V}}\right)$ for the different models. \label{fig:ionhist}}
\end{figure}

In the upper panel of Figure~\ref{fig:ionhist}, we show the volume
weighted ionization fraction $1 - x_{\mathrm{V}}$ as a function of
time for three of our high-resolution radiative transfer
calculations. We see a rapid, exponential rise at high redshift, and
an approach to unity at around $z\simeq 6-8$. The end of reionization
and the rapid phase transition to a reionized universe is better
visible in displaying the neutral fraction $x_{\mathrm{V}}$, which is
also included in the figure. For comparison, we also show
observational constraints derived by \citet{Fan2006a} from quasar
absorption lines (symbols with error bars). Especially our model
V1\_1024\_M1 reproduces the suggested end of reionization in these data
quite well, although the residual neutral fraction comes out slightly
low.

In the lower panel of Figure~\ref{fig:ionhist}, we consider the
evolution of the ratio between mass- and volume-weighted ionization
fractions,
$\left (1 - x_{\mathrm{M}} \right)/\left(1 - x_{\mathrm{V}} \right)$.
This quantity is the average density of the ionized hydrogen compared to
the average hydrogen density of the full box:
\begin{align}
  \frac{1 - x_{\mathrm{M}}}{1 - x_{\mathrm{V}}} = \frac{V_{ \mathrm{tot} } }{M_{\mathrm{tot}}}
  \frac{\left (1 - x_{\mathrm{M}} \right) M_{\mathrm{tot}}}{\left(1 - x_{\mathrm{V}} \right) V_{\mathrm{tot}}} = \frac{1}{\langle
    \rho \rangle_{\mathrm{tot}}}
  \frac{M_{\mathrm{ionized}}}{V_{\mathrm{ionized}}} = \frac{\langle
    \rho \rangle_{\mathrm{ionized}}}{\langle \rho
    \rangle_{\mathrm{tot}}}.
\end{align}
This ratio stays at or above unity for all time, which can be interpreted
as a signature of an inside-out character of reionization
\citep{Iliev2006}. Overdense environments around our sources ionize first,
so that the ionized volume is always overdense on average. Interestingly,
the evolution of the mean overdensity of ionized regions with time also
differs for our different escape fraction models. The run assuming a
constant escape fraction model starts reionization later but then
progresses somewhat more rapidly. This model maintains the highest value
of $\left (1 - x_{\mathrm{M}} \right)/\left(1 - x_{\mathrm{V}} \right)$
for most of the simulated timespan. Here reionization is particularly
biased to overdense regions and is stuck there for a comparatively long
time, until the final reionization transition occurs on a short timescale
and the IGM at mean density is ionized as well. Interestingly, even though
the variable escape fraction models show some variety in the time of the
onset of reionization and the remaining neutral fraction, the evolution of
$\left (1 - x_{\mathrm{M}} \right)/\left(1 - x_{\mathrm{V}} \right)$ is
still rather similar among these models. They begin reionization earlier
and thus generally have low mean overdensities of the ionized volume at
any given time.

\begin{figure}
\begin{center}
  \setlength{\unitlength}{1cm}
\hspace*{-0.4cm}\resizebox{9.3cm}{!}{\includegraphics{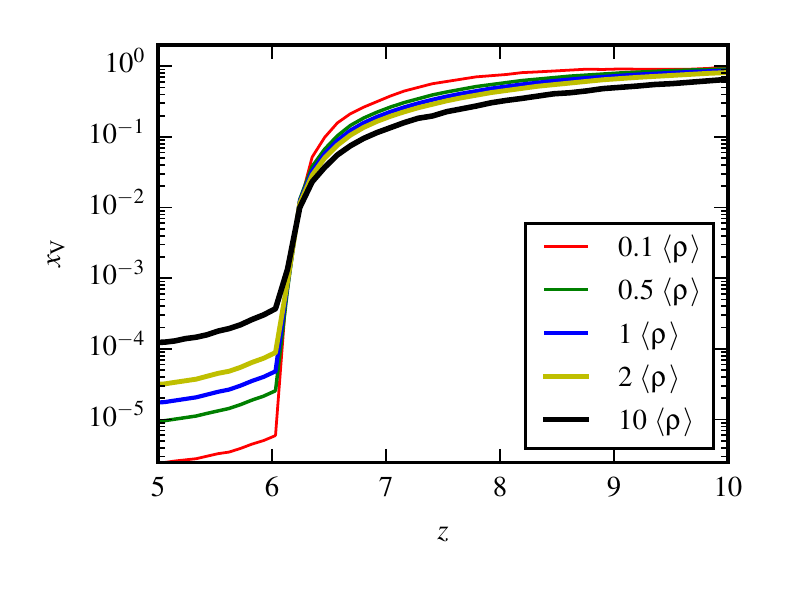}}\vspace*{-0.45cm}%
\caption{Time evolution of the neutral hydrogen fraction at given
  gas overdensities. Each of the lines shows the result for a small density
  range around a nominal comoving density threshold, as labeled, for
  our model V1\_1024\_M1. We see that the ionization of dense gas begins
  earlier, but this gas also ends up with a higher residual neutral fraction once reionization is completed.
  \label{fig:ion_density}}
\end{center}
\end{figure}

That the character of the reionization process is best described as an
inside out transition can be seen in more detail in
Figure~\ref{fig:ion_density}, where the time evolution of the average
neutral fraction is shown for regions of a fixed given
overdensity. Highly overdense regions start to become ionized quite
early on, assisted by collisional ionization in virialized halos. As a
result, the reionization process for high density regions is also
considerably less sudden than for lower density gas. After
reionization is essentially completed, the behavior however reverses;
now overdense regions show on average a final ionization degree that
is considerably lower than for lower density regions. This can be
understood as a result of the higher recombination rate in the denser
regions, shifting the equilibrium value of the ionized fraction in a
given UV background. Generally speaking, we find that denser regions
start to ionize earlier but keep a higher neutral fraction than
underdense regions.

Reionization is clearly not an instantaneous transition but requires a
certain amount of time. It is hence interesting to characterize the epoch
of reionization not only with a single redshift but also to ask how
long the duration to a reionized universe takes. To this end we define
the duration $\Delta z$ as the interval in redshift space during which
the volume-weighted neutral fraction drops from $80 \% $ to $20 \% $.
Our fiducial model V1\_1024\_M1 leads to an extent of
$\Delta z = 2.28$ for the epoch of reionization, corresponding to a
time\-span of $\Delta t = 251.2\, \mathrm{Myr}$ for this period. The
V5\_1024\_M1 model shows a slightly longer duration of reionization
with $\Delta z = 2.61$ and $\Delta t = 341.3\, \mathrm{Myr}$.
Reionization lasts only over a span of $\Delta z = 1.13$ or
$\Delta t = 190.0\, \mathrm{Myr}$ in our C2\_1024\_M1 model. Thus our
models with a variable escape fraction show a more extended epoch of
reionization compared to the model with a constant escape fraction.
In the models with a variable escape fraction, the ionizing luminosity
is initially higher, but once most of the ionizing luminosity has
become available, the variable escape fraction gradually begins to
limit the amount of escaping UV radiation, resulting in a more
prolonged epoch of reionization.

\begin{figure}
\begin{center}
  \setlength{\unitlength}{1cm} \hspace*{-0.4cm}\resizebox{9.3cm}{!}{\includegraphics{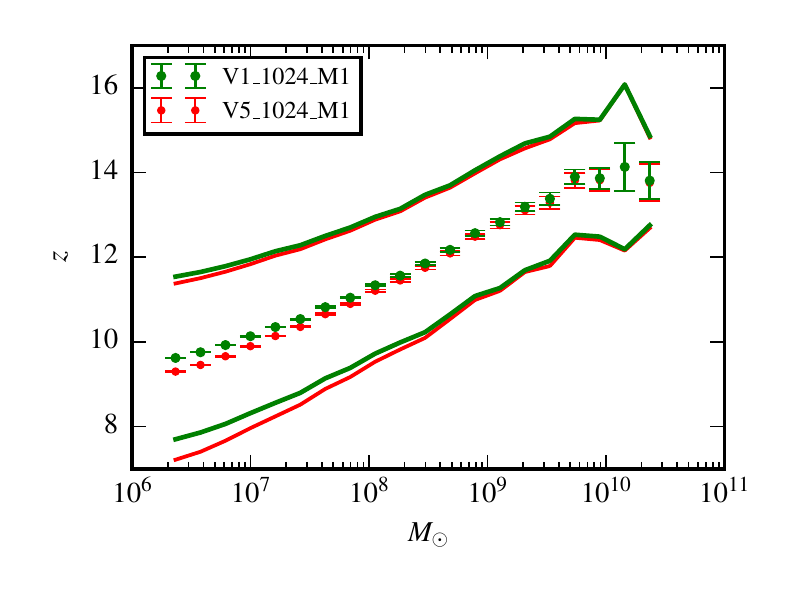}}\vspace*{-0.45cm}%
  \caption{Mean reionization redshifts of the immediate surroundings of
  galaxies as a function of their $z=6$ stellar masses for our highest
  resolution run of the V1 and V5 models. The symbols give the means in
  each mass bin, with the error bars showing the statistical error of the
  mean. The solid lines illustrate the $\pm1\sigma$ variance in each mass
  bin. \label{fig:ion_time_galaxies}}
\end{center}
\end{figure}

One can also ask whether galaxies of different stellar masses are all
ionized at the same time, or whether there are significant systematic
trends of the mean reionization epoch as a function of galaxy size. To
this end, we have considered the sample of all $z=6$ galaxies in Illustris
and then checked at what redshift the average ionized fraction in a sphere
of radius $150\,{\rm kpc}$ around them (taking their $z=6$ positions)
reached 50\% for the first time. The results of this analysis are shown in
Figure~\ref{fig:ion_time_galaxies}, binned as a function of stellar mass.
There is a clear trend for an earlier reionization around more massive
galaxies. Interestingly, the spread in the reionization times slightly
increases towards smaller stellar masses, indicating that dwarf galaxies
are expected to show larger diversity in their reionization histories.
Also, these low mass galaxies are more sensitive to the adopted
parametrization of the escape fraction and tend to reionize later in our
$V5$ model.

\subsection{UV background}

\begin{figure}
\begin{center}
\setlength{\unitlength}{1cm}
\hspace*{-0.5cm}\resizebox{9.3cm}{!}{\includegraphics{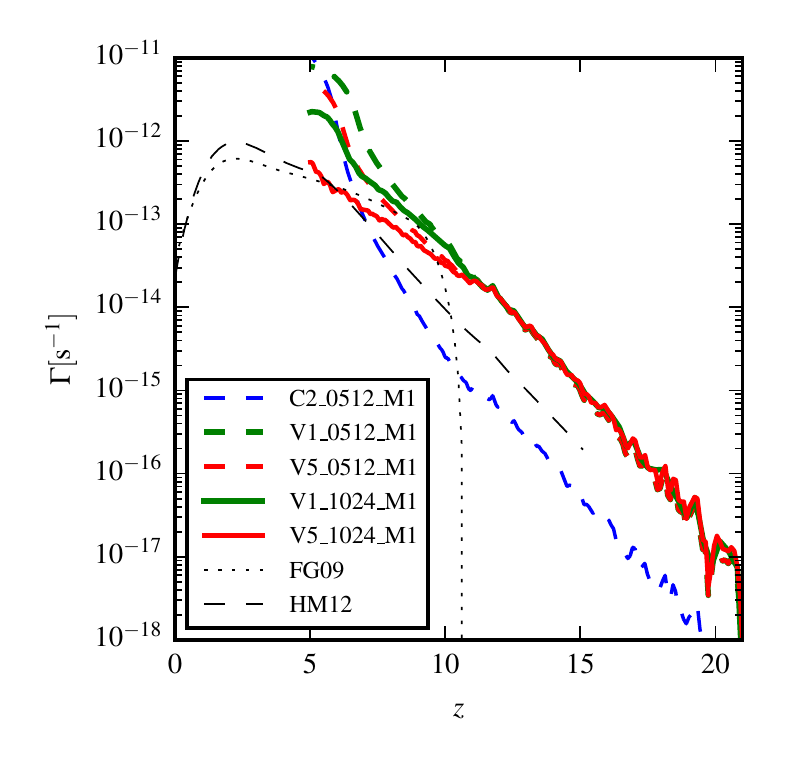}}\vspace*{-0.5cm}%
\caption{Average photoionization rate $\Gamma$ as function of redshift
  for models calculated with different escape fractions and grid
  resolutions.  For comparison, we also include two widely used
  theoretical models for the UV background evolution: FG09 \citep[][we
  show the updated version from Dec. 2011]{Faucher-Giguere2009} and
  HM12 \citep{Haardt2012}. The V5\_1024\_M1 model is in particularly
  good agreement with FG09, at least for $z < 10$. 
\label{fig:uvb}}
\end{center}
\end{figure}

In Figure~\ref{fig:uvb}, we consider the time evolution of the volume
averaged photoionization rate for models calculated with different escape
fractions and grid resolutions, and compare to different models in the
literature for the evolution of the cosmic UV background
\citep{Haardt2012, Faucher-Giguere2009}. At high redshift, the background
builds up exponentially with redshift, similar to the growth of the volume
weighted ionization fraction, with an overall amplitude that varies with
the escape fraction model. Our variable escape fraction models and our
fixed escape fraction model bracket the scenario of HM12. Interestingly,
our scenario V5\_1024\_M1 follows the model of \citet{Faucher-Giguere2009}
very closely, except that after reionization is completed, our
calculations tend to overshot the model predictions. We note that a very
similar behaviour is also seen in the recent radiative transfer
simulations of \citet{Pawlik2015}, where this effect is even more
pronounced. A more realistic variable escape fraction model than the
rather simple parametrization employed might help resolving this issue.

\subsection{Optical depth $\tau$}

\begin{figure}
\begin{center}
\setlength{\unitlength}{1cm}
\hspace*{-0.6cm}\resizebox{9.55cm}{!}{\includegraphics{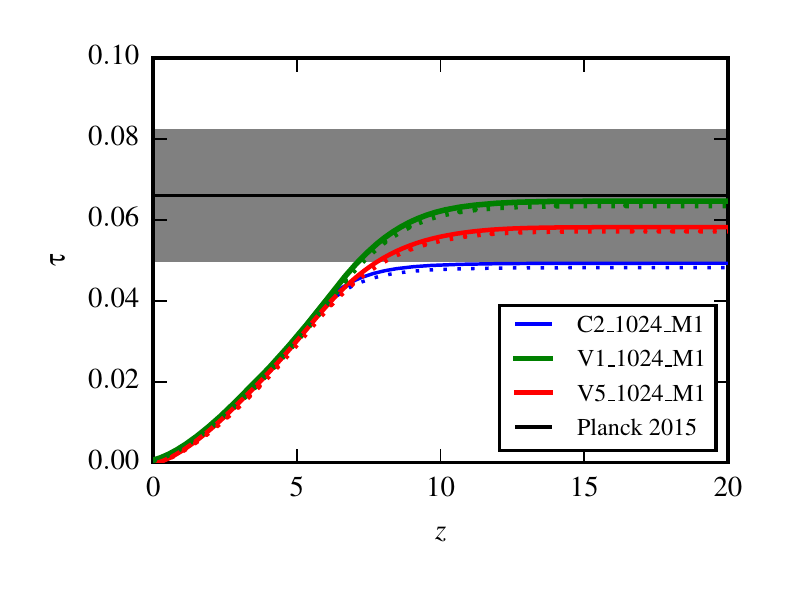}}\vspace*{-0.5cm}%
\caption{Cumulative optical depth for Thompson scattering on free
  electrons, integrated out to the redshift specified on the
  horizontal axis. Solid lines include electrons from doubly ionized
  helium (assuming that they contribute for $z<3$), while dotted lines
  assume hydrogen and one helium electron only. The horizontal line
  with $\pm 1\sigma$ uncertainty region (shaded) marks the newest 2015
  constraints $\tau = 0.066 \pm 0.016 $ by the
  \citet{PlanckCollaboration2015}.  Out fiducial model V1\_1024\_M1 is
  in very good agreement with optical depth inferred from these
  precision measurements of the CMB. Our other models lie slightly
  lower, however their value is still consistent. Interestingly, the
  previous determination by Planck based on their first 2013 data
  analysis had given a considerably higher value for $\tau$. The
  tension with this result is now resolved.\label{fig:opticaldepth}}
\end{center}
\end{figure}

Starting with the onset of reionization, CMB photons will Thomson
scatter off the free electrons again. This effect can be quantified by
measuring the cumulative optical depth $\tau$ seen by CMB photons
along their path towards us.  This optical depth is given by
\begin{align}
\tau = c\, \sigma_{\mathrm{th}} \int^0_{z_0} n_e(z) \frac{\mathrm{d} t}{\mathrm{d}z} \mathrm{d}z,
\end{align}
where $\sigma_{\mathrm{th}}$ is the Thomson cross section, and $n_e$
the number density of free electrons. In this work, we only consider
hydrogen reionization for simplicity. Given that the first ionization
potential of helium is very close to the ionization potential of
hydrogen, we assume that HeII is created in proportion to HII. The
free electron density is then given by
\begin{align}
    n_e = \left[1+\frac{1-X}{4X}\right] \left(1 - x_{\mathrm{M}} \right) n_{\rm H} \approx 1.079\, \left(1 - x_{\mathrm{M}} \right) n_{\rm H},
\end{align}     
where $X=0.76$ is the hydrogen mass fraction. At late times, helium
reionization will eventually be completed, increasing the optical
depth slightly compared to the above estimate. Adopting the common
assumption that helium becomes doubly ionized at $z=3$
\citep[e.g.][]{Iliev2005}, the free electron density increases by
$\Delta n_e = (1-X)/(4X) \left(1 - x_{\mathrm{M}} \right) n_{\rm H}
\simeq 0.079\, \left(1 - x_{\mathrm{M}} \right) n_{\rm H}$
compared to the estimate above, and hence the optical is enlarged by
\begin{align}
    \Delta \tau = 
c\, \sigma_{\mathrm{th}} \int^0_{3} \Delta n_e(z) \frac{\mathrm{d} t}{\mathrm{d}z} \mathrm{d}z
= 0.0011,
\end{align}     
which is a very small correction given the other uncertainties.

The optical depth $\tau$ for a specific reionization history can be
converted into an effective reionization redshift $z_{\mathrm{reion}}$
assuming a fiducial scenario in which the reionization transition is
instantaneous at this epoch. The latest {\small WMAP9} results find a
best-fit value of $\tau = 0.088\pm0.013$, corresponding to
$z_{\mathrm{reion}} = 10.5\pm 1.1$ \citep{Hinshaw2013}, quite a bit lower
than the value of $\tau = 0.17\pm 0.08$ {\small WMAP1} had initially
estimated. The results of the {\small PLANCK} mission in its 2013 data
release \citep{Planck2013CosmoParams} favour a very similar, slightly
larger value for the optical depth, $\tau = 0.089\pm0.032$, corresponding
to an even earlier reionization redshift of $z_{\mathrm{reion}} = 10.8$.
However, the latest {\small PLANCK} data release of 2015
\citep{PlanckCollaboration2015} prefers a much lower optical depth of
$\tau = 0.066\pm 0.016$ and a corresponding redshift of reionization of
$z_{\mathrm{reion}} = 8.8^{+1.7}_{-1.4}$. A similar low optical
depth of $\tau = 0.063 \pm 0.013$ has been found in
\citet{Finkelstein2014} based on a UV luminosity function derived from
Hubble Ultra Deep Field and Hubble Frontier Field data.

In Figure~\ref{fig:opticaldepth}, we show the optical depth of our
reionization simulations as a function of the integration redshift $z$
for three of our models. The most recent 2015 constraint from {\small
  PLANCK} is shown as a horizontal line, together with the
$\pm 1\sigma$ uncertainty region (shaded). Our models with a variable
escape fraction are comfortably compatible within the error bars with
the 2015 Planck results. Our fiducial model V1\_1024\_M1 predicts an
optical depth of $\tau = 0.065$ which is in very good agreement with
the most recent Planck 2015 data. However all of our other models
prefer the low side of the range determined by Planck. Still, it is
very promising that the former tension between galaxy formation
simulations and optical depth inferred from CMB measurements seems
nearly resolved with the 2015 Planck data. A similar finding has been
reported in \citet{Robertson2015} based on Hubble observations of
distant galaxies. Allowing for additional high redshift star formation
could easily close the small remaining gap if needed, but we note that
constraints from galaxy formation disfavour this solution. For
example, Illustris already tends to overshoot estimates for the
stellar mass function of small galaxies, at late and early times alike
\citep{Vogelsberger2014b, Genel2014}. Resolving these problems seems
to call for reduced high redshift star formation and not the opposite,
highlighting the difficulty to reconcile high optical depths values
from CMB experiments with detailed galaxy formation models.

\subsection{Bubble size statistics}

\begin{figure}
\begin{center}
\setlength{\unitlength}{1cm}
\hspace*{-0.5cm}\resizebox{9.3cm}{!}{\includegraphics{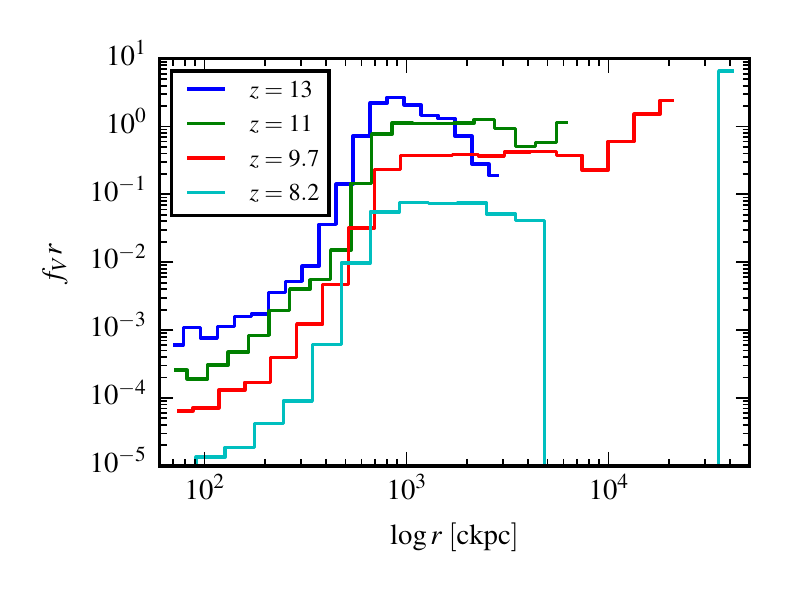}}%
\vspace*{-0.5cm}
\caption{Distribution of the characteristics radius 
of ionized regions at four different redshifts, in our fiducial model
V1\_1024\_M1.  The integral over the distributions is normalized to unity for each of the measurements, and a horizontal line corresponds to
equal volume fraction per logarithmic size interval.
 Initially, small bubbles dominate but over time the distribution
 shifts to ever larger bubbles until only one large region dominates.
 \label{fig:regions}}
\end{center}
\end{figure}

\begin{figure}
\begin{center}
\setlength{\unitlength}{1cm}
\hspace*{-0.5cm}\resizebox{9.3cm}{!}{\includegraphics{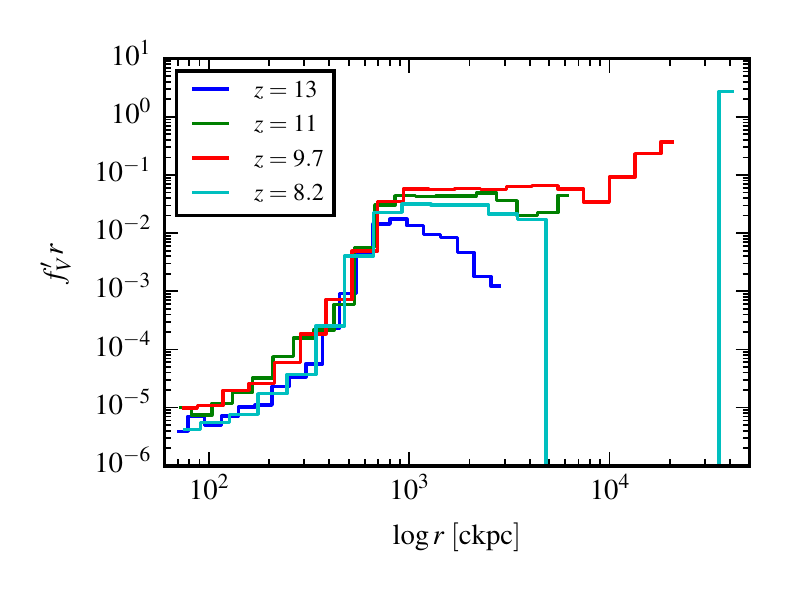}}
\vspace*{-0.6cm}
\caption{The same information as in Fig.~\ref{fig:regions}, except
  with a different normalization. Here the integral is normalized by
  the constant comoving box volume so that the area under the  
  distributions gives the total ionized volume fraction. This is
  informative because the relative constancy of the size distributions
  for small bubble sizes suggests that the growing bubbles 
  are replenished by new small bubbles just at the right rate to
  achieve this balance. Interestingly, the volume occupied by small bubbles
  stays hence roughly the same for the whole duration of reionization, 
  even slightly beyond bubble percolation where 
  a dominating single large ionized region forms.
  \label{fig:regions2}}
\end{center}
\end{figure}

Mapping the epoch of reionization more directly than possible thus
far, for example through 21cm imaging, is an exciting observational
prospect. Once this becomes possible with future radio telescopes such
as the SKA, quantitative measures of the geometry of the ionized
regions, such as their topology, promise to be a powerful probe of
theoretical models for reionization. Radiation transfer models like
those calculated herein are the method of choice to make the required
detailed predictions about how these morphological measures evolve in
time. To illustrate this, we here compute a few basic statistical
measures that quantify the number and size of ionized regions as a
function of time, which may also serve as a useful comparison against
other theoretical reionization models.

We start by tagging cells as ionized if their ionization fraction exceeds
$1 - x_{\mathrm{M}} > 0.5$. Based on the resulting grid of binary values,
ionized regions are then identified using a friends-of-friends algorithm
that links adjacent ionized cells \citep{Iliev2006,Chardin2012}. We let cells belong
to the same group if they share at least one corner, or in other words, if
any of the $26$ neighbours of an ionized cell is also ionized, both cells
are put into the same group. For each ionized island identified in this
way, we compute an effective radius as $r = \left[{3}/\left({4 \pi} V
\right) \right]^{1/3}$, where $V$ is the cumulative volume of the cells
making up the group. Finally, we consider the distribution function $f_V$
of the ionized volume fraction contained in regions of a given bubble
radius.

In Figure~\ref{fig:regions}, we show the resulting distribution
function when the convention of \cite{Zahn2007} is followed and the
distribution is normalized such that
\begin{align}
\int f_V r  \, \mathrm{d}\!\log r = 1,
\end{align}
i.e.~we only consider the ionized volume of the box. The results show
that at early times, when only a small fraction of the volume is
ionized, the ionized volume is comprised of disjoint regions of
characteristic size $r \simeq 2\,{\rm cMpc}$. While reionization
progresses, ever larger bubbles appear, with a flat distribution as a
function of size, i.e.~roughly the same amount of ionized volume is
contained per logarithmic interval in bubble size, up to bubble sizes
of the order of $r \simeq 20\,{\rm cMpc}$. Eventually, the bubbles start to
percolate and one dominating region containing a substantial fraction
of the simulation volume is formed ($z=8.2$). There is then still a
population of smaller ionized regions left, with a constant volume
fraction per unit $\log r$ over a dynamic range of about $\sim 5$ in
size.

An alternative normalization for the size distribution is used in
\cite{Gnedin2014b}, who consider the whole box thus that integrating
over $f_V$ gives the ionized volume fraction:
\begin{align}
\int f'_V\, r \, \mathrm{d}\!\log r = 1 - x_{\mathrm{V}}.
\end{align}
It is instructive to plot the corresponding distributions also with
this normalization, which is shown in Figure~\ref{fig:regions2}. Now
the area under the distribution function grows with redshift,
reflecting the increase of the ionized volume fraction. Interestingly,
we see in this representation that the bubble size distribution is
fairly constant with time, especially for the small bubble sizes. Even
though these bubbles grow individually in size with time, the fact
that the abundance of bubbles of a given size stays approximately
constant in time (once the first bubbles of this size have formed),
suggests that small bubbles are reformed just at the right rate to
compensate for the loss of bubbles of a given size due to the growth
or coalescence of bubbles.

\begin{figure}
\begin{center}
\setlength{\unitlength}{1cm}
\hspace*{-0.3cm}\resizebox{9cm}{!}{\includegraphics{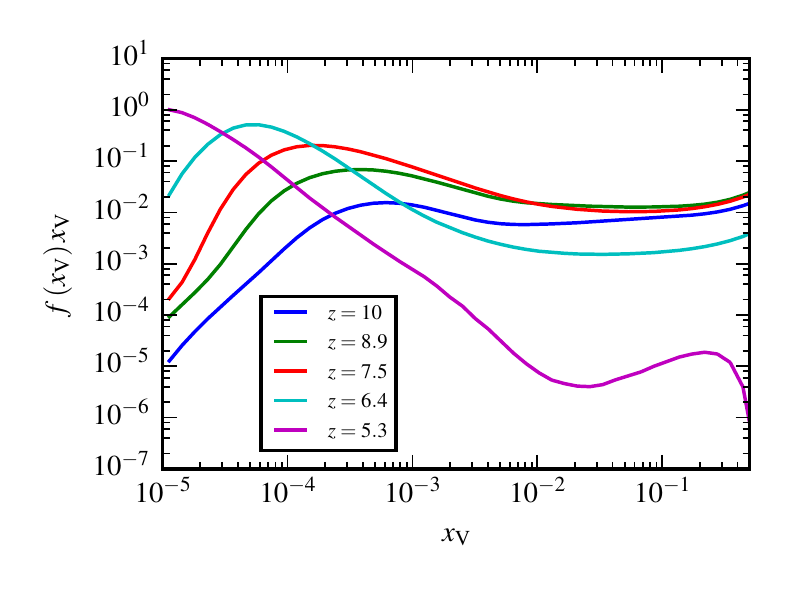}}\\%
\vspace*{-0.5cm}\hspace*{-0.3cm}\resizebox{9cm}{!}{\includegraphics{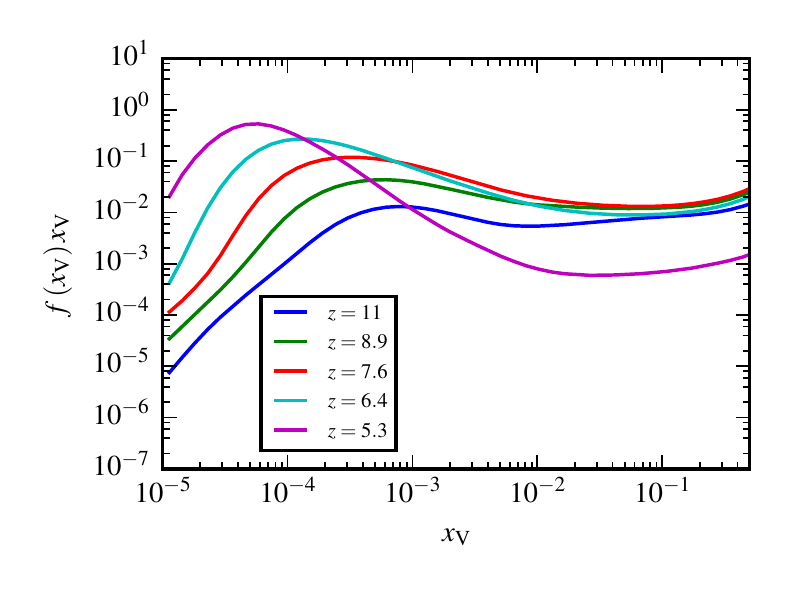}}%
\vspace*{-0.35cm}
\caption{Degree of ionization level PDFs at different redshifts.  The
  top panel shows the results for the V1\_1024\_M1 model, whereas
  the bottom panel is for the V5\_1024\_M1 model. Compare to
  Fig.~\ref{fig:uvb}, where the V1 run shows a strong upturn in the UV
  flux, which is here expressed as a shift of ionized cells down to
  $n_{\mathrm{HI}} = 10^{-5}$ at around $z=5.3$, which does not happen
  in the V5 run.
  \label{fig:pdf}}
\end{center}
\end{figure}

\subsection{Distribution function of neutral volume fraction}

In Figure~\ref{fig:pdf}, we show the distribution function
$f(x_{\mathrm{V}}) = {\rm d}V / {\rm d}x_{\mathrm{V}}$ of the volume
fraction that is found at a given neutral fraction. We compare our two
different variable reionization scenarios, in the form of V1\_1024\_M1
(top panel) and V5\_1024\_M1 (bottom panel), and give results for
different redshifts in each case. Note that we plot
$x_{\mathrm{V}}\,f(x_{\mathrm{V}})$ on the vertical axis versus the log of
$x_{\mathrm{V}}$, i.e.~the area under the curves is proportional to the
volume fraction at the corresponding range of neutral fractions.

Progress in reionization is associated with a large increase in the volume
fraction found at low neutral fractions, as is of course expected.
Interestingly, the differential distribution of volume at a given neutral
fraction is however fairly broad while reionization is not completed, with
a peak at a characteristic neutral fraction that shifts progressively to
lower values. For example, at redshifts $z\simeq 11$, most of the volume
is either still neutral or at a neutral fraction of $x_{\mathrm{V}}\sim
10^{-3}$. At around redshift $z\sim 7$, the characteristic neutral
fraction where most of the volume is found has dropped to
$x_{\mathrm{V}}\sim 10^{-4}$. Finally, post reionization, the two models
start to differ more prominently. Here V1\_1024\_M1 shows a low neutral
fraction of $x_{\mathrm{V}}\sim 10^{-5}$ or lower for most of its volume,
which corresponds also to the strong rise in the predicted UV background
seen in this model in Figure~\ref{fig:uvb}. In contrast, the model
V5\_1024\_M1, which shows good agreement with the UV background model of
\citet{Faucher-Giguere2009} at this epoch, yields a markedly different
distribution of the neutral fraction, with most of the volume having
neutral fractions around $x_{\mathrm{V}}\simeq 5\times 10^{-5}$.

\section{Caveats and Discussion}  \label{sec:numerics}
\label{sec:discussion}

The accuracy and reliability of our results are influenced by many
numerical aspects as well as physical uncertainties. In the following
we discuss a number of these aspects, focusing primarily on those
pertaining to the radiative transfer modelling itself. We note however
that there are in principle additional uncertainties related, for
example, to the treatment of star formation and the associated
feedback processes in the underlying Illustris simulation, or to the
cosmological background model that we use. These are arguably
subdominant compared to the uncertainties related to the reionization
calculation itself (such as escape fraction, radiative transfer
solver, etc.), and in any case are beyond the scope of this paper
\citep[a discussion of the uncertainties in the galaxy formation model
can be found in][]{Vogelsberger2014b}.

\subsection{Reionization feedback}

Due to the fact that we simulate reionization only in post-processing,
any back reaction onto the gas due to photoionization heating and
potentially radiation pressure is not taken into account
self-consistently. The Illustris simulation assumes a uniform global
UV background, hence the {\em average} back reaction on star formation
due to photo-ionization is approximately accounted for, but any local
modulation of the corresponding effects is of course ignored. This
limitation could only be overcome by dynamically coupling the
radiative transfer solver to a hydrodynamical code and doing full
radiation-hydrodynamics simulations of galaxy formation. Recently,
impressive progress has been made in this direction
\citep{Gnedin2014a,Gnedin2014b,Pawlik2015}, but the achieved
cosmological volumes are still severely limited due to the demanding
computational cost of radiative transfer, and in general, these
calculations have not been evolved to redshift $z=0$, and thus it is
unclear whether they are successfully reproducing the observed galaxy
population.

Besides photo-heating, the ionizing radiation of young stars could
also exert significant feedback effects through radiation pressure,
particularly in dusty gas where infrared radiation may be trapped
\citep{Murray2010, Hopkins2012, Agertz2013}. However, the
effectiveness of this mechanism is debated, with a number of recent
studies arguing that photo-heating is likely the dominant feedback
channel on the scale of galaxies, with radiation pressure being
comparatively unimportant \citep{Sales2014, Rosdahl2015}. We thus
consider the omission of radiation pressure effects in our
reionization calculations to be comparatively unimportant.

\subsection{Moment-based versus cone-based RT method}

Our radiative transfer implementation supports two different transport
methods, allowing us to compare them against each other with no
changes in any other aspect of the modelling. In the case of the
cone-based method, we have to store and process 48 radiation intensity
fields, one for each advection direction. On the other hand, the
moment-based M1 scheme only requires 4 fields, one for the photon
number density and 3 for the flux vector. This difference makes the
cone-based advection scheme much more expensive in terms of
computational cost as well as in terms of (GPU) memory requirements.

\begin{figure}
\begin{center}
\setlength{\unitlength}{1cm}
\hspace*{-0.3cm}\resizebox{9cm}{!}{\includegraphics{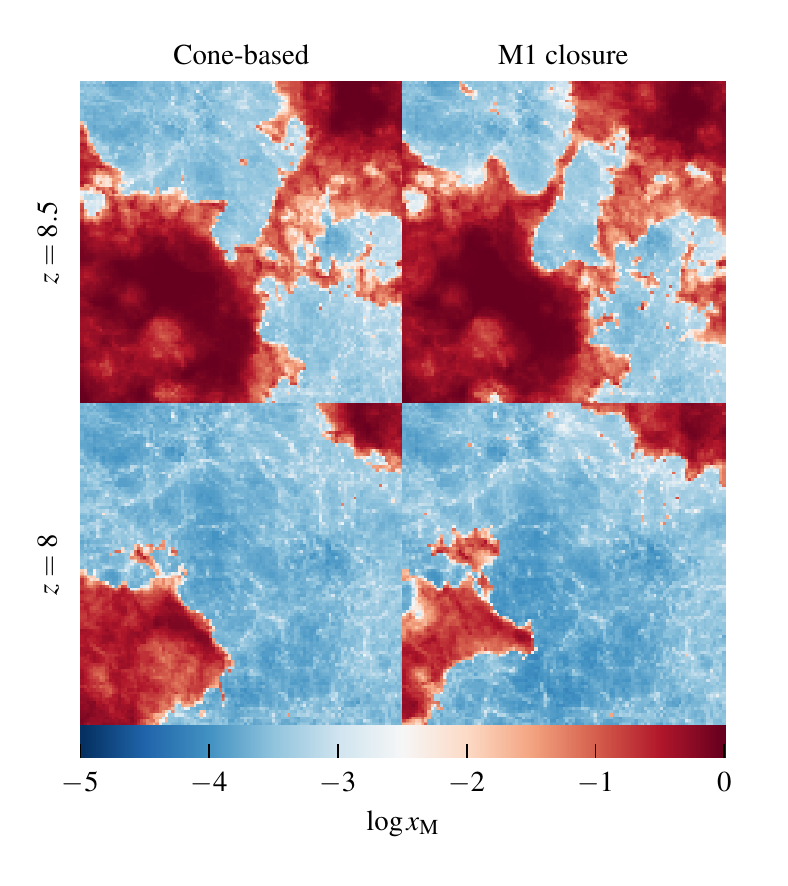}}\vspace*{-0.2cm}%
\caption{Visual comparison of our two different radiation advection
methods at two instances in time. All panels show a thin slice through the
box with a side length of $21.3\, \textrm{cMpc}$. The top row compares the
neutral hydrogen fraction at $z=8.5$ for the cone-based method
(left) and the moment-based method with M1 closure (right). Similarly, the
bottom row shows this comparison for redshift $z=8$. \label{fig:galcomp}}
\end{center}
\end{figure}

If we compare the ionization histories predicted by these different
radiative transfer methods in terms of the ionized volume fraction, no
appreciable differences are detected. In fact, the agreement of the
evolution of the ionized volume fractions is so good that we refrain
from showing the corresponding comparison in a dedicated plot. But
this consistency is perhaps not too surprising. The ionization history
mainly depends on the source population that injects ionizing photons,
as well as on the density evolution of the gas, as the latter
sensitively determines the recombination rate. Given that the photon
injection rates and the density structure are exactly equal in our
comparison, and given the fact that both radiative transfer algorithms
are manifestly photon conserving, any difference between the
cone-based and the M1-closure methods can only be induced by
differences in the photon transport directions. These are apparently
subtle enough that they do not matter much for global statistics of
the reionization transition.

However, despite this good agreement in global averages, the two methods
still show differences in the detailed morphology of the ionized bubbles
when examined in detail. In Figure~\ref{fig:galcomp}, we compare the
morphology of the ionized regions around a typical galaxy at redshifts
$z=8$ and $z=8.5$. While there is clearly a great deal of similarity, the
detailed locations of the ionization fronts differ substantially,
highlighting that the radiative transfer solvers do not behave identically
after all. This is also borne out by a higher order quantitative
comparison of the neutral hydrogen fraction fields predicted by
the two methods. This can for example be done by computing the
mass-weighted standard deviation of the difference between the neutral
hydrogen densities obtained by our two radiative transfer schemes:
\begin{align}
   v = \mathrm{var} \left ( \rho \left( x_{\mathrm{M1}} - x_{\mathrm{CONE}}\right )\right) / \left<\rho\right>. 
\end{align}
For our V1\_0512 models, we find that this quantity rises with
decreasing redshift until a maximum of $v = 0.2$ is reached at
$z=7.5$. Afterwards, the full volume is quickly reionized and the
variance of the difference field rapidly declines again. We note that
the cone-based method should be the more accurate approach in this
comparison, as it can avoid certain inaccuracies of the M1 approach,
in particular when the ionization bubbles of two or more sources
overlap.

\subsection{How accurate is the reduced speed of light approximation?}
\label{sec:rsl}
As long as the medium is dense enough, the propagation speed of the
ionization fronts is determined by the rate at which new ionizing
photons arrive at the edge of neutral gas, and not how fast they get
there.  This motivates the idea of the so-called reduced speed of
light approximation \citep{Gnedin2001, Aubert2008}, in which the
physical value of $c$ is artificially reduced.  The computational
advantage of a reduced speed of light is that a much larger Courant
time step is allowed in schemes where photon transport is followed
with explicit time integration.  \citet{Rosdahl2013} report that the
reduced speed of light approximation describes the solution of
Str\"omgren sphere well after an effective crossing time
$t_{\mathrm{cross}} = r_{\mathrm{S}}/c$, where $r_{\mathrm{S}}$ is the
radius of the corresponding Str\"omgren sphere and $c$ the (reduced)
speed of light. Before $t_{\mathrm{cross}} = r_{\mathrm{S}}/c$,
however, the numerical solution necessarily always falls behind the
correct one.  Considering the relevant time scales that have to be
resolved in cosmic reionization, this yields a criterium for the
maximum allowed reduction of the speed of light. The conclusion of
\citet{Rosdahl2013} is that there is not much room for applying the
reduced speed of light approximation if accurate reionization
simulations of the IGM are desired.

\begin{figure}
\begin{center}
\setlength{\unitlength}{1cm}
\hspace*{-0.4cm}\resizebox{9.2cm}{!}{\includegraphics{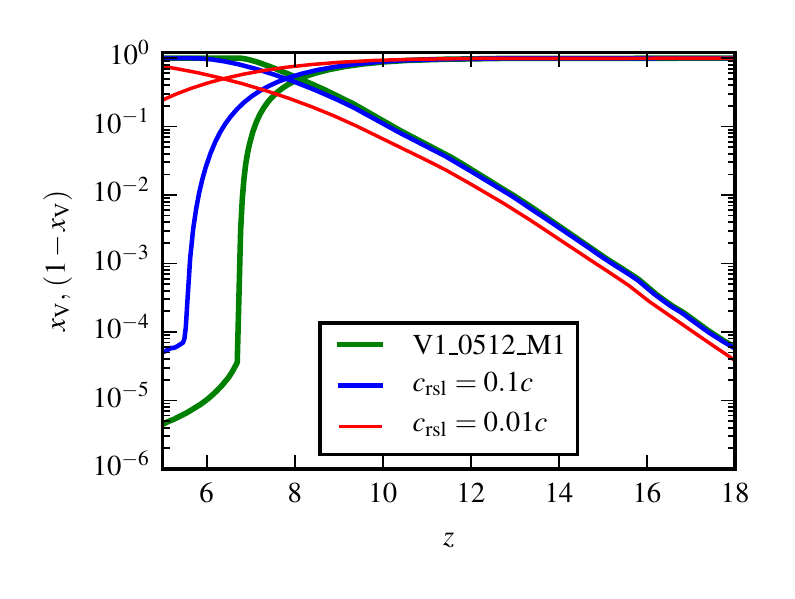}}\vspace*{-0.4cm}%
\caption{Impact of the reduced speed of light approximation on the evolution of the neutral and 
ionized volume fractions. Reducing the speed of light by a factor of 10 (blue line) or 100 (red line)
relative to our default calculation with a physical speed of light
(green line) causes a later reionization of the universe. 
\label{fig:rsl}}
\end{center}
\end{figure}

It is interesting to use our independent radiative transfer code to
check this assessment.  In Figure~\ref{fig:rsl}, we show the impact of
the reduced speed of light on the obtained reionization history, based
on a reduction of the physical speed of light by a factor of 10 or 100,
respectively. Consistently with the findings of \citet{Rosdahl2013},
the reduction of the speed of light leads to a significant delay in
the resulting epoch of reionization, amounting to
$\Delta z \sim 1-1.5$ for the factor of 10 reduction, and much larger
for a factor of 100. The size of this error unfortunately implies that
this numerical trick can induce unacceptably large distortions in the
reionization predictions, hence we have refrained from using it
throughout the study.

\subsection{Clumping factors}

\begin{figure}
\begin{center}
\setlength{\unitlength}{1cm}
\hspace*{-0.4cm}\resizebox{9.2cm}{!}{\includegraphics{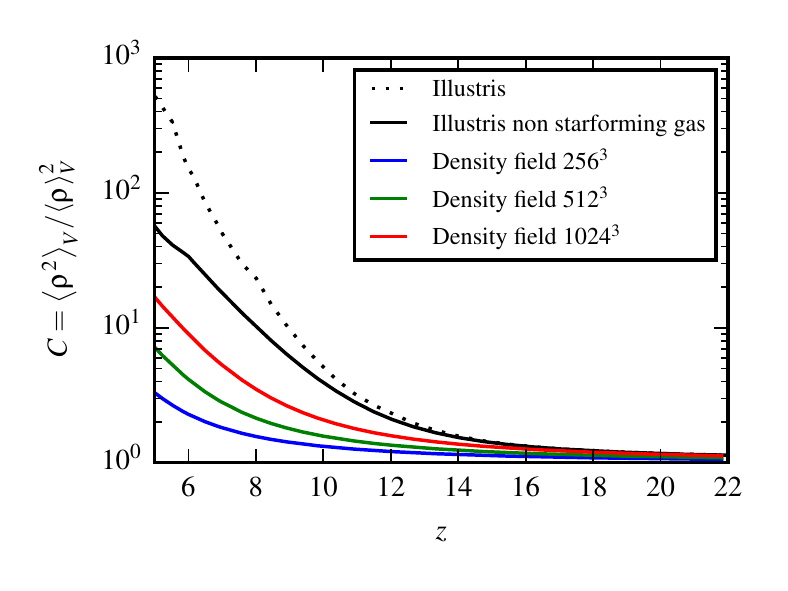}}\vspace*{-0.4cm}%
\caption{Clumping factor of the gas computed in different ways.  The
  upper solid line shows the intrinsic clumping factor of all non
  star-forming gas (with density below the star formation threshold),
  based on the Voronoi tessellation of the Illustris simulation volume.  The
  dotted line gives the intrinsic clumping factor of all the gas in
  the simulation.  For comparison, we also show the clumping factor of
  all the non star-forming gas when this is obtained for different
  mesh resolutions after mapping the simulation volume to a grid with
  fixed spatial resolution.  \label{fig:ClumpingConvergence}}
\end{center}
\end{figure}

\begin{figure*}
\begin{center}
\setlength{\unitlength}{1cm}
\hspace*{-0.4cm}\resizebox{9.2cm}{!}{\includegraphics{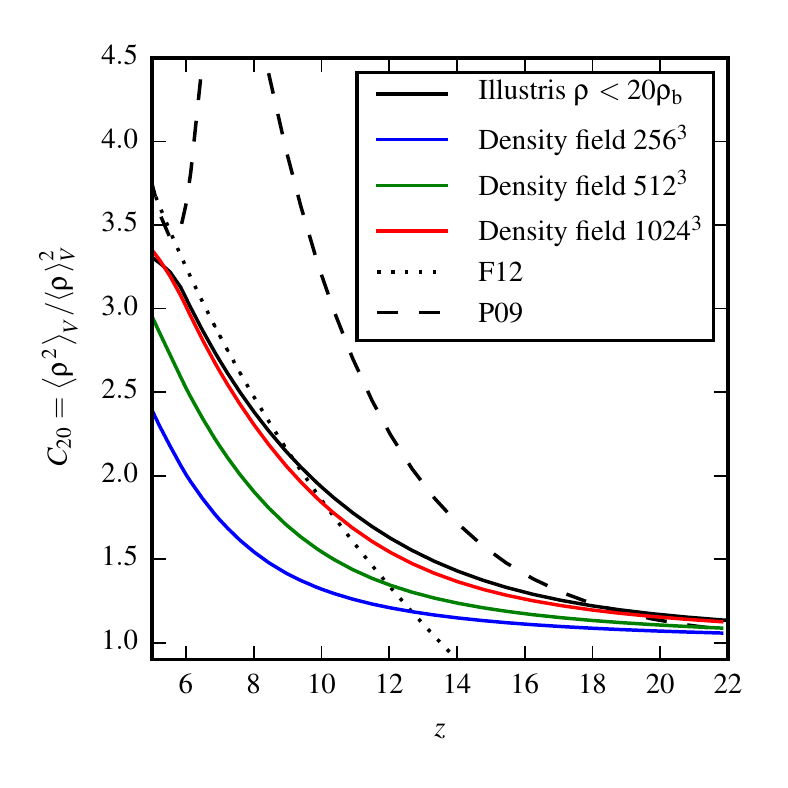}}%
\hspace*{-0.4cm}\resizebox{9.2cm}{!}{\includegraphics{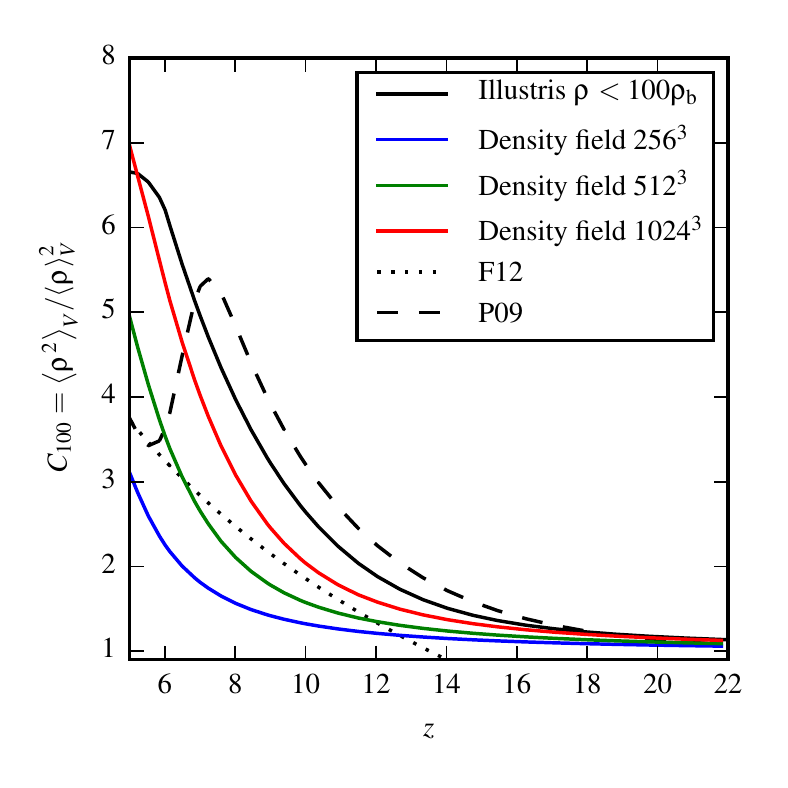}}\vspace*{-0.4cm}%
\caption{Gas clumping factors measured for gas below a certain
characteristic gas overdensity. The left-hand panel shows $C_{20}$, with the
solid black line giving the result for the actual Voronoi tessellation
used in the Illustris simulation for the hydrodynamical calculations. The
other solid lines give the corresponding result for the binned density
field when grids with resolution from $256^3$ to $1024^3$ are used, as
labeled. The right-hand panel gives the same results for $C_{100}$, where
instead a density threshold of 100 times the mean baryonic density is
adopted. For reference, we also show fitting models by
\citet[][F12]{Finlator2012} and \citep[][P09]{Pawlik2009}, which give the
clumping factor of ionized gas or gas below a overdensity threshold of 100
(their $z \sim 7.5$ reionization case), respectively.
\label{fig:clumping_20}}
\end{center}
\end{figure*}

The degree of gas clumping is a critical factor in models of
reionization, as it directly determines the recombination rate and
hence the amount of photons required to reionize the universe and to
keep it ionized. Full hydrodynamical simulations of galaxy formation
are a particularly powerful tool to realistically predict the
non-linear density structure of the gas, and hence to quantify the gas
clumping.  In Figure~\ref{fig:ClumpingConvergence}, we show the {\em
  full} clumping factor $C$ of all the gas, defined in the standard
way as
\begin{align}
C = \frac{\left < \rho^2 \right>}{\left < \rho \right >^2},
\end{align}
where the averages are volume averages for the full simulation box.
The black lines represent the clumping factor obtained directly
from the Voronoi cells of the underlying Illustris simulation, which
is hence accounting for all structure resolved by the more than 6
billion cells of the simulation. We give results for the complete
density field (dotted line) as well as for cells constrained to not
lie on the effective equation of state of star forming gas (solid
black line). The former result includes the collapsed gas that corresponds
to the ISM and is star-forming, whereas in the latter this phase is
excluded. The clumping factor is obviously much higher when this
star-forming gas is included, but as the sub-grid model used by
Illustris glosses over the true multi-phase nature of the ISM, the
resulting clumping factor for all the gas is still an
underestimate. However, we are here really only interested in the
non-starforming gas, because the recombinations and absorptions inside
the star forming regions are collectively accounted for by the escape
fraction, which in part may be viewed as parameterizing our ignorance
of the detailed gas structure on ISM scales. 

The solid red, green and blue lines in Fig.~\ref{fig:ClumpingConvergence}
show the clumping factor of all gas after binning it on radiation transfer
grids with different resolution. Clearly, even for the $1024^3$ grid we
loose almost a factor of two in the total clumping due to the smoothing of
this grid. However, as cosmic reionization is a volume filling process and
the densest gas occupies only a tiny fraction of the value, it makes more
sense to refer the escape fraction to a somewhat lower density threshold
than the star-formation threshold. We are hence really interested in the
clumping factor of gas up to some limited overdensity value, for example
up to 20 or 100 times the mean baryonic density. Since both of these
fiducial values have been used in the literature, we show in
Figure~\ref{fig:clumping_20} our results for $C_{20}$ and $C_{100}$ (in
the left- and right-hand panels, respectively), where only gas cells which have
a density of at most $20 \times \rho_{\mathrm{b}}$ or $100 \times
\rho_{\mathrm{b}}$ have been included, respectively, with
$\rho_{\mathrm{b}}= \Omega_{\rm b} \rho_{\mathrm{crit}}$ denoting the mean
baryon density. As before, we show results for the underlying Voronoi
tessellation as well as reionization grids between $256^3$ and $1024^3$
resolutions. Note that in these plots a linear scale for the clumping
factor has been used. 

Our high resolution radiative transfer grid underpredicts the
$C_{\rm 20}$ clumping a bit due to the smoothing effects of the
binning, but the effect is minor. The situation is a bit worse if one
also wants to get the correct clumping factor for gas up to an
overdensity of 100. Here some of this additional clumping is not
resolved by the $1024^3$ grid, as the results in the right-hand panel of
Fig.~\ref{fig:clumping_20} show. Given the trend with increasing
resolution, using a $2048^3$ grid instead (which we unfortunately
cannot carry out due to memory constraints on the GPU system we have
presently access to) should however be able to fully recover the
$C_{100}$ clumping of this gas. As we discuss in more detail below,
this resolution problem for the $C_{100}$ quantity affects cosmic
reionization however only mildly and is hence comparatively benign.

It is interesting to compare our clumping factors with those inferred from
other works. \citet{Finlator2012} have pointed out that different
definitions of the clumping factor can produce substantial differences in
the results. It is thus important to base any such comparison on the same
definition, which sometimes corresponds to considering the clumping of all
the gas below a certain density threshold (to separate collapsed and
diffuse gas), or to restricting the evaluation of the clumping factor to
ionized gas. Note that the latter depends both on the detailed
reionization model and the gas distribution. To get a sense of how well
the gas distributions compare, it is thus arguably best to compare the
total gas clumping factor. For the Illustris simulation at $z=8$, we
measure for the clumping of the non-starforming gas $10.2$, slightly
higher than the value reported by \citet{Finlator2012}. However, our value
is significantly higher than the total gas clumping factor of $C\sim 3$ at
$z=8$ reported by \citet{Jeeson-Daniel2014}, which makes it considerably
easier to achieve reionization in their model. When the baryon density is
restricted to lie below an overdensity of 100, we find a clumping factor
of about 4 at $z=8$, somewhat larger than what was found in
\citet{Pawlik2009} for their models reionizing at or before $z = 9$, but a
bit lower than their model reionizing at $z=7.5$. For a yet lower
overdensity threshold of 20, \citet{Wise2014} report a value of 6.5, which
is above our measurement of $\simeq 2.4$ for $C_{20}$ at redshift $z=8$.
This likely reflects the higher mass resolution of their simulation, which
has a boxsize of just $1\,{\rm Mpc}$, but it could also be affected by the
different feedback models in the two simulations.

\subsection{Spatial resolution and convergence}

Because the recombination rate depends nonlinearly on the density in a
cell, the smoothing of density fluctuations (for example as a result
of binning) causes an underestimate of recombination events and hence
biases reionization towards higher redshift.  Our results for the
clumping factor indicate that our radiative transfer calculations
clearly suffer from this effect to some degree. However, it is not
obvious whether the size of the bias is quantitatively significant in
the end, because the clumping of the volume-filling gas (which has
comparatively low overdensity) is captured well by our high-resolution
grid. If reionization would mostly occur `outside-in', with low
density regions ionized early, one may hope that this is already
sufficient for allowing converged predictions of the reionization
redshift even if density peaks are washed out. However, given that our
results have confirmed that dense regions tend to be ionized
earlier, this may largely be wishful thinking.

\begin{figure}
\begin{center}
\setlength{\unitlength}{1cm}
\hspace*{-0.4cm}\resizebox{9.3cm}{!}{\includegraphics{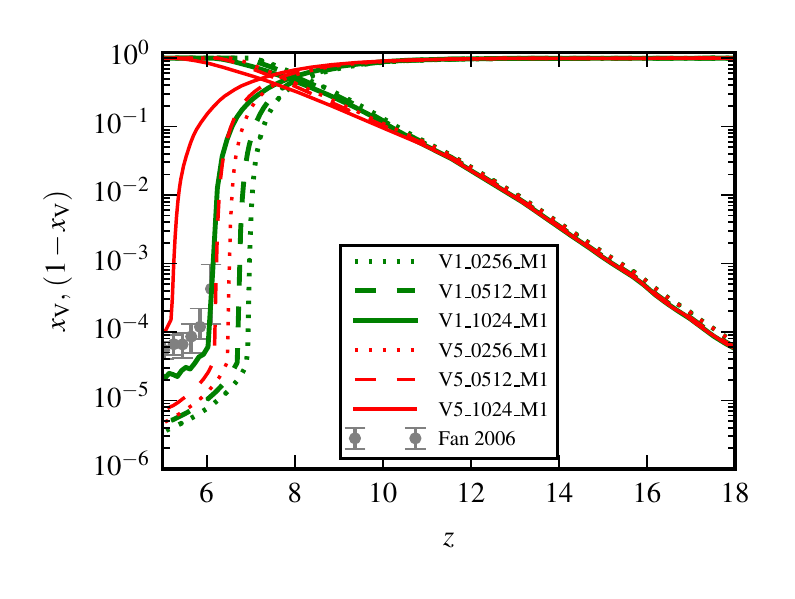}}\vspace*{-0.4cm}%
\caption{Convergence study for the neutral and ionized volume fractions for our reionization simulations based on the M1 method. The  green and red lines are for the two different variable escape fraction models we considered. There are significant residual trends with resolution due to the smoothing effects a coarse grid has on the clumpy gas distribution. As a result, the highest resolution simulation tends to reionize slightly later than predicted by calculations at lower resolution.
  \label{fig:convergence}}
\end{center}
\end{figure}

Indeed, this is borne out by our convergence tests for the
reionization history of models V1 and V5 shown in
Figure~\ref{fig:convergence}. Evidently, as the resolution of the grid
for the radiative transfer simulation is increased, reionization
happens progressively somewhat later, as a result of the smaller
degree of suppression of the true underlying clumpiness of the
gas. This prevents us from achieving a formal numerical convergence
for our reionization histories. We note however that full radiation
hydrodynamics simulations will counter this drift by physically
inducing a reduction of the clumpiness of the gas \citep{Pawlik2009},
due to the photo-heating and the resulting pressure smoothing. This
effect is not included in our simulations prior to redshift $z=6$,
when reionization happens due to the externally imposed UV
background. We thus expect that our $1024^3$ and $512^3$ grids may
well bracket the true behavior of the clumping in a self-consistent
simulation with full radiation hydrodynamics. This then also means
that a $2048^3$ calculation {\em without} taking this effect into
account may well produce a less accurate result than the $1024^3$ grid
we used.

\section{Conclusions}  \label{sec:conclusion}

In this work, we considered only hydrogen reionization and focused on
ordinary high-redshift star formation as the primary source of
ionizing photons. Other populations may in principal contribute
to reionization, in particular primordial population-III stars, AGNs,
or exotic sources such as annihilating dark matter. While Pop-III
stars may be important for the onset of reionization, most estimates
for their relative contribution to global star formation suggest that
neglecting them for reionization is justified. Still, accounting for
them in future models would be clearly desirable, if only for
completeness. Our neglect of AGN radiation for hydrogen reionization
is however quite well justified because their ionizing luminosity is
overwhelmed by star formation at high redshift. AGNs become important
at intermediate redshifts, however, where they likely play an
important role in HeII reionization.

Even with these simplifying assumptions, calculating the ionizing flux
that becomes available for reionization is affected by a substantial
number of uncertainties. This includes the stellar population
synthesis model we have used, and in particular, the adopted stellar
initial mass function, where we used a Chabrier IMF and assumed that
there are no significant IMF variations as a function of environment.
Another major uncertainty lies in the escape fraction, which is
physically uncertain and is in large part a phenomenological
parameter in our models, absorbing uncertainties due to the treatment
of dense gas and the limited spatial resolution. Finally, there are
also numerical limitations related to the radiative transfer solver,
the finite angular and spatial resolution employed in the radiative
transfer, and the lack of self-consistently accounting for local
feedback effects by the radiation field.

Fortunately some of these uncertainties can be greatly reduced by matching
key observables such as the total optical depth for electron scattering or
the amplitude of the metagalactic ionizing UV background after completion
of reionization \citep[e.g.][]{Faucher-Giguere2008a,Faucher-Giguere2008b}.
Our radiative transfer models for Illustris then basically test whether
cosmic reionization does occur for reasonable assumptions about the escape
fraction, based on a galaxy population that yields a successful
description of a slew of other observational data both at high and low
redshift. To the extent this works, it provides reassurance for the
cosmological consistency of our galaxy formation and reionization models,
and it shows that they are physically meaningful. Importantly, they can
hence be used to learn more about the reionization transition itself. Our
main findings can be summarized as follows:

\begin{enumerate}
\item The star formation history predicted by the Illustris simulation
  combined with its high-resolution gas density field allows cosmic
  reionization with an optical depth of $\tau = 0.065$, consistent with
  the latest Planck 2015 results as well as constraints from
  high-redshift quasars. This relies on ordinary stellar populations
  only, but requires optimistic assumptions for a high escape fraction
  at high redshift.
\item Previous tensions between the high optical depth favoured by CMB
  results and the low level of high redshift star formation required
  by successful galaxy formation models are thus essentially resolved
  with Planck 2015.
\item Using a suitable variable escape fraction model, we can
  approximately reproduce the expected UV background after
  reionization is completed, but most of our models tend to then
  overshoot the ionization rate and yield a slightly lower neutral
  fraction than inferred from quasar absorption lines. A fine-tuned
  model may plausibly yield an improved match.
\item Reionization proceeds inside out in our model, with overdense
  regions being ionized earlier on average than lower density regions.
\item The size distribution of ionized regions shows a remarkable
  constancy with time for small bubble sizes, suggesting that during
  reionization new small bubbles are formed roughly at the rate at
  which they are removed by size growth or coalescence with other
  regions. The characteristic size of bubbles grows with time, but
  there is a fairly flat distribution of sizes between
  $r\sim 2\,{\rm cMpc}$ and $r \sim 20 \,{\rm cMpc}$ with about equal
  volume fraction per logarithmic size interval just before
  reionization is completed.
\item The duration of the reionization transition varies with the
  escape fraction model, and we find transition times for a drop of
  the neutral fraction from 80 to 20\% between 190 and 340 Myr,
  depending on the model.
\item The distribution of volume with respect to neutral fraction is
  quite broad but shows a peak that progresses to ever smaller neutral
  fraction with time. For models that successfully match the UV
  background constraints after reionization, the characteristic neutral
  fraction is $5\times 10^{-5}$, with the lowest amount of volume
  found at neutral fractions of $2\times 10^{-2}$.
\end{enumerate}

In future work based on our methodology it would be particularly
interesting to study HeII reionization. Due to the two ionization
levels of helium with different ionization thresholds, the
computational cost rises by at least a factor of about three, as more
spectral bins have to be tracked instead of just one for
hydrogen. Additionally, a much longer physical time down to $z\sim 3$
has to be followed. This would still be possible in postprocessing
with a future Illustris type simulation with a larger box size,
containing an evolving AGN population including contributions from the
brightest objects. Simulating this much physical time is extremely
challenging for direct radiation-hydrodynamics simulations, much more
so than hydrogen reionization simulations that can be stopped at
$z\simeq 6$. This makes postprocessing approaches the only practical
radiative transfer method to study HeII reionization in the near term.

\section*{Acknowledgements}
We thank the anonymous referee for constructive criticism and valuable
suggestions that helped to improve the paper. We also thank R\"udiger Pakmor
and Ewald Puchwein for helpful discussions. AB and VS acknowledge support
by the European Research Council under ERC-StG EXAGAL-308037. AB
acknowledges support by the IMPRS for Astronomy \& Cosmic Physics at the
University of Heidelberg. LH acknowledges support from NASA grant
NNX12AC67G and NSF grant AST-1312095. SG acknowledges support for programme
number HST-HF2-51341.001-A that was provided by NASA through a Hubble
Fellowship grant from the Space Telescope Science Institute, which is
operated by the Association of Universities for Research in Astronomy,
Incorporated, under NASA contract NAS5-26555. We gratefully thank for
GPU-accelerated computer time on the `Milky Way' GPU cluster of the DFG
Research Centre SFB-881 `The Milky Way System', and on `Hydra' at the
Max-Planck Computing Centre RZG in Garching.

\bibliographystyle{mnras}
\bibliography{paper}

\begin{thebibliography}{}
\makeatletter
\relax
\def\mn@urlcharsother{\let\do\@makeother \do\$\do\&\do\#\do\^\do\_\do\%\do\~}
\def\mn@doi{\begingroup\mn@urlcharsother \@ifnextchar [ {\mn@doi@}
  {\mn@doi@[]}}
\def\mn@doi@[#1]#2{\def\@tempa{#1}\ifx\@tempa\@empty \href
  {http://dx.doi.org/#2} {doi:#2}\else \href {http://dx.doi.org/#2} {#1}\fi
  \endgroup}
\def\mn@eprint#1#2{\mn@eprint@#1:#2::\@nil}
\def\mn@eprint@arXiv#1{\href {http://arxiv.org/abs/#1} {{\tt arXiv:#1}}}
\def\mn@eprint@dblp#1{\href {http://dblp.uni-trier.de/rec/bibtex/#1.xml}
  {dblp:#1}}
\def\mn@eprint@#1:#2:#3:#4\@nil{\def\@tempa {#1}\def\@tempb {#2}\def\@tempc
  {#3}\ifx \@tempc \@empty \let \@tempc \@tempb \let \@tempb \@tempa \fi \ifx
  \@tempb \@empty \def\@tempb {arXiv}\fi \@ifundefined
  {mn@eprint@\@tempb}{\@tempb:\@tempc}{\expandafter \expandafter \csname
  mn@eprint@\@tempb\endcsname \expandafter{\@tempc}}}

\bibitem[\protect\citeauthoryear{{Abel} \& {Wandelt}}{{Abel} \&
  {Wandelt}}{2002}]{Abel2002}
{Abel} T.,  {Wandelt} B.~D.,  2002, \mn@doi [\mnras]
  {10.1046/j.1365-8711.2002.05206.x}, \href
  {http://adsabs.harvard.edu/abs/2002MNRAS.330L..53A} {330, L53}

\bibitem[\protect\citeauthoryear{{Abel}, {Norman}  \& {Madau}}{{Abel}
  et~al.}{1999}]{Abel1999}
{Abel} T.,  {Norman} M.~L.,   {Madau} P.,  1999, \mn@doi [\apj]
  {10.1086/307739}, \href {http://adsabs.harvard.edu/abs/1999ApJ...523...66A}
  {523, 66}

\bibitem[\protect\citeauthoryear{{Agertz}, {Kravtsov}, {Leitner}  \&
  {Gnedin}}{{Agertz} et~al.}{2013}]{Agertz2013}
{Agertz} O.,  {Kravtsov} A.~V.,  {Leitner} S.~N.,   {Gnedin} N.~Y.,  2013,
  \mn@doi [\apj] {10.1088/0004-637X/770/1/25}, \href
  {http://adsabs.harvard.edu/abs/2013ApJ...770...25A} {770, 25}

\bibitem[\protect\citeauthoryear{{Ahn}, {Iliev}, {Shapiro}, {Mellema}, {Koda}
  \& {Mao}}{{Ahn} et~al.}{2012}]{Ahn2012}
{Ahn} K.,  {Iliev} I.~T.,  {Shapiro} P.~R.,  {Mellema} G.,  {Koda} J.,   {Mao}
  Y.,  2012, \mn@doi [\apjl] {10.1088/2041-8205/756/1/L16}, \href
  {http://adsabs.harvard.edu/abs/2012ApJ...756L..16A} {756, L16}

\bibitem[\protect\citeauthoryear{{Alvarez} \& {Abel}}{{Alvarez} \&
  {Abel}}{2007}]{Alvarez2007}
{Alvarez} M.~A.,  {Abel} T.,  2007, \mn@doi [\mnras]
  {10.1111/j.1745-3933.2007.00342.x}, \href
  {http://adsabs.harvard.edu/abs/2007MNRAS.380L..30A} {380, L30}

\bibitem[\protect\citeauthoryear{{Aubert} \& {Teyssier}}{{Aubert} \&
  {Teyssier}}{2008}]{Aubert2008}
{Aubert} D.,  {Teyssier} R.,  2008, \mn@doi [\mnras]
  {10.1111/j.1365-2966.2008.13223.x}, \href
  {http://adsabs.harvard.edu/abs/2008MNRAS.387..295A} {387, 295}

\bibitem[\protect\citeauthoryear{{Aubert} \& {Teyssier}}{{Aubert} \&
  {Teyssier}}{2010}]{Aubert2010}
{Aubert} D.,  {Teyssier} R.,  2010, \mn@doi [\apj]
  {10.1088/0004-637X/724/1/244}, \href
  {http://adsabs.harvard.edu/abs/2010ApJ...724..244A} {724, 244}

\bibitem[\protect\citeauthoryear{{Barkana} \& {Loeb}}{{Barkana} \&
  {Loeb}}{2001}]{Barkana2001}
{Barkana} R.,  {Loeb} A.,  2001, \mn@doi [\physrep]
  {10.1016/S0370-1573(01)00019-9}, \href
  {http://adsabs.harvard.edu/abs/2001PhR...349..125B} {349, 125}

\bibitem[\protect\citeauthoryear{{Battaglia}, {Trac}, {Cen}  \&
  {Loeb}}{{Battaglia} et~al.}{2013}]{Battaglia2013}
{Battaglia} N.,  {Trac} H.,  {Cen} R.,   {Loeb} A.,  2013, \mn@doi [\apj]
  {10.1088/0004-637X/776/2/81}, \href
  {http://adsabs.harvard.edu/abs/2013ApJ...776...81B} {776, 81}

\bibitem[\protect\citeauthoryear{{Bennett} et~al.,}{{Bennett}
  et~al.}{2013}]{Bennett2013}
{Bennett} C.~L.,  et~al., 2013, \mn@doi [\apjs] {10.1088/0067-0049/208/2/20},
  \href {http://adsabs.harvard.edu/abs/2013ApJS..208...20B} {208, 20}

\bibitem[\protect\citeauthoryear{{Bird}, {Haehnelt}, {Neeleman}, {Genel},
  {Vogelsberger}  \& {Hernquist}}{{Bird} et~al.}{2015}]{Bird2014}
{Bird} S.,  {Haehnelt} M.,  {Neeleman} M.,  {Genel} S.,  {Vogelsberger} M.,
  {Hernquist} L.,  2015, \mn@doi [\mnras] {10.1093/mnras/stu2542}, \href
  {http://adsabs.harvard.edu/abs/2015MNRAS.447.1834B} {447, 1834}

\bibitem[\protect\citeauthoryear{{Bouwens} et~al.,}{{Bouwens}
  et~al.}{2011}]{Bouwens2011}
{Bouwens} R.~J.,  et~al., 2011, \mn@doi [\apj] {10.1088/0004-637X/737/2/90},
  \href {http://adsabs.harvard.edu/abs/2011ApJ...737...90B} {737, 90}

\bibitem[\protect\citeauthoryear{{Caruana}, {Bunker}, {Wilkins}, {Stanway},
  {Lorenzoni}, {Jarvis}  \& {Ebert}}{{Caruana} et~al.}{2014}]{Caruana2014}
{Caruana} J.,  {Bunker} A.~J.,  {Wilkins} S.~M.,  {Stanway} E.~R.,  {Lorenzoni}
  S.,  {Jarvis} M.~J.,   {Ebert} H.,  2014, \mn@doi [\mnras]
  {10.1093/mnras/stu1341}, \href
  {http://adsabs.harvard.edu/abs/2014MNRAS.443.2831C} {443, 2831}

\bibitem[\protect\citeauthoryear{{Chardin}, {Aubert}  \& {Ocvirk}}{{Chardin}
  et~al.}{2012}]{Chardin2012}
{Chardin} J.,  {Aubert} D.,   {Ocvirk} P.,  2012, \mn@doi [\aap]
  {10.1051/0004-6361/201219992}, \href
  {http://adsabs.harvard.edu/abs/2012A%26A...548A...9C} {548, A9}

\bibitem[\protect\citeauthoryear{{Ciardi}, {Ferrara}  \& {White}}{{Ciardi}
  et~al.}{2003}]{Ciardi2003}
{Ciardi} B.,  {Ferrara} A.,   {White} S.~D.~M.,  2003, \mn@doi [\mnras]
  {10.1046/j.1365-8711.2003.06976.x}, \href
  {http://adsabs.harvard.edu/abs/2003MNRAS.344L...7C} {344, L7}

\bibitem[\protect\citeauthoryear{{Croft} \& {Altay}}{{Croft} \&
  {Altay}}{2008}]{Croft2008}
{Croft} R.~A.~C.,  {Altay} G.,  2008, \mn@doi [\mnras]
  {10.1111/j.1365-2966.2008.13513.x}, \href
  {http://adsabs.harvard.edu/abs/2008MNRAS.388.1501C} {388, 1501}

\bibitem[\protect\citeauthoryear{{Davis}, {Stone}  \& {Jiang}}{{Davis}
  et~al.}{2012}]{Davis2012}
{Davis} S.~W.,  {Stone} J.~M.,   {Jiang} Y.-F.,  2012, \mn@doi [\apjs]
  {10.1088/0067-0049/199/1/9}, \href
  {http://adsabs.harvard.edu/abs/2012ApJS..199....9D} {199, 9}

\bibitem[\protect\citeauthoryear{{Dillon} et~al.,}{{Dillon}
  et~al.}{2014}]{Dillon2014}
{Dillon} J.~S.,  et~al., 2014, \mn@doi [\prd] {10.1103/PhysRevD.89.023002},
  \href {http://adsabs.harvard.edu/abs/2014PhRvD..89b3002D} {89, 023002}

\bibitem[\protect\citeauthoryear{{Ellis} et~al.,}{{Ellis}
  et~al.}{2013}]{Ellis2013}
{Ellis} R.~S.,  et~al., 2013, \mn@doi [\apjl] {10.1088/2041-8205/763/1/L7},
  \href {http://adsabs.harvard.edu/abs/2013ApJ...763L...7E} {763, L7}

\bibitem[\protect\citeauthoryear{{Fan}, {Carilli}  \& {Keating}}{{Fan}
  et~al.}{2006a}]{Fan2006}
{Fan} X.,  {Carilli} C.~L.,   {Keating} B.,  2006a, \mn@doi [\araa]
  {10.1146/annurev.astro.44.051905.092514}, \href
  {http://adsabs.harvard.edu/abs/2006ARA%26A..44..415F} {44, 415}

\bibitem[\protect\citeauthoryear{{Fan} et~al.,}{{Fan} et~al.}{2006b}]{Fan2006a}
{Fan} X.,  et~al., 2006b, \mn@doi [\aj] {10.1086/504836}, \href
  {http://adsabs.harvard.edu/abs/2006AJ....132..117F} {132, 117}

\bibitem[\protect\citeauthoryear{{Faucher-Gigu{\`e}re}, {Lidz}, {Hernquist}  \&
  {Zaldarriaga}}{{Faucher-Gigu{\`e}re} et~al.}{2008a}]{Faucher-Giguere2008a}
{Faucher-Gigu{\`e}re} C.-A.,  {Lidz} A.,  {Hernquist} L.,   {Zaldarriaga} M.,
  2008a, \mn@doi [\apjl] {10.1086/590409}, \href
  {http://adsabs.harvard.edu/abs/2008ApJ...682L...9F} {682, L9}

\bibitem[\protect\citeauthoryear{{Faucher-Gigu{\`e}re}, {Lidz}, {Hernquist}  \&
  {Zaldarriaga}}{{Faucher-Gigu{\`e}re} et~al.}{2008b}]{Faucher-Giguere2008b}
{Faucher-Gigu{\`e}re} C.-A.,  {Lidz} A.,  {Hernquist} L.,   {Zaldarriaga} M.,
  2008b, \mn@doi [\apj] {10.1086/592289}, \href
  {http://adsabs.harvard.edu/abs/2008ApJ...688...85F} {688, 85}

\bibitem[\protect\citeauthoryear{{Faucher-Gigu{\`e}re}, {Lidz}, {Zaldarriaga}
  \& {Hernquist}}{{Faucher-Gigu{\`e}re} et~al.}{2009}]{Faucher-Giguere2009}
{Faucher-Gigu{\`e}re} C.-A.,  {Lidz} A.,  {Zaldarriaga} M.,   {Hernquist} L.,
  2009, \mn@doi [\apj] {10.1088/0004-637X/703/2/1416}, \href
  {http://adsabs.harvard.edu/abs/2009ApJ...703.1416F} {703, 1416}

\bibitem[\protect\citeauthoryear{{Finkelstein} et~al.,}{{Finkelstein}
  et~al.}{2014}]{Finkelstein2014}
{Finkelstein} S.~L.,  et~al., 2014, preprint, \href
  {http://adsabs.harvard.edu/abs/2014arXiv1410.5439F} {} (\mn@eprint {arXiv}
  {1410.5439})

\bibitem[\protect\citeauthoryear{{Finlator}, {{\"O}zel}  \&
  {Dav{\'e}}}{{Finlator} et~al.}{2009}]{Finlator2009}
{Finlator} K.,  {{\"O}zel} F.,   {Dav{\'e}} R.,  2009, \mn@doi [\mnras]
  {10.1111/j.1365-2966.2008.14190.x}, \href
  {http://adsabs.harvard.edu/abs/2009MNRAS.393.1090F} {393, 1090}

\bibitem[\protect\citeauthoryear{{Finlator}, {Oh}, {{\"O}zel}  \&
  {Dav{\'e}}}{{Finlator} et~al.}{2012}]{Finlator2012}
{Finlator} K.,  {Oh} S.~P.,  {{\"O}zel} F.,   {Dav{\'e}} R.,  2012, \mn@doi
  [\mnras] {10.1111/j.1365-2966.2012.22114.x}, \href
  {http://adsabs.harvard.edu/abs/2012MNRAS.427.2464F} {427, 2464}

\bibitem[\protect\citeauthoryear{{Furlanetto}, {Zaldarriaga}  \&
  {Hernquist}}{{Furlanetto} et~al.}{2004}]{Furlanetto2004}
{Furlanetto} S.~R.,  {Zaldarriaga} M.,   {Hernquist} L.,  2004, \mn@doi [\apj]
  {10.1086/423025}, \href {http://adsabs.harvard.edu/abs/2004ApJ...613....1F}
  {613, 1}

\bibitem[\protect\citeauthoryear{{Genel} et~al.,}{{Genel}
  et~al.}{2014}]{Genel2014}
{Genel} S.,  et~al., 2014, \mn@doi [\mnras] {10.1093/mnras/stu1654}, \href
  {http://adsabs.harvard.edu/abs/2014MNRAS.445..175G} {445, 175}

\bibitem[\protect\citeauthoryear{{Gnedin}}{{Gnedin}}{2000}]{Gnedin2000}
{Gnedin} N.~Y.,  2000, \mn@doi [\apj] {10.1086/308876}, \href
  {http://adsabs.harvard.edu/abs/2000ApJ...535..530G} {535, 530}

\bibitem[\protect\citeauthoryear{{Gnedin}}{{Gnedin}}{2014}]{Gnedin2014a}
{Gnedin} N.~Y.,  2014, \mn@doi [\apj] {10.1088/0004-637X/793/1/29}, \href
  {http://adsabs.harvard.edu/abs/2014ApJ...793...29G} {793, 29}

\bibitem[\protect\citeauthoryear{{Gnedin} \& {Abel}}{{Gnedin} \&
  {Abel}}{2001}]{Gnedin2001}
{Gnedin} N.~Y.,  {Abel} T.,  2001, \mn@doi [\na]
  {10.1016/S1384-1076(01)00068-9}, \href
  {http://adsabs.harvard.edu/abs/2001NewA....6..437G} {6, 437}

\bibitem[\protect\citeauthoryear{{Gnedin} \& {Kaurov}}{{Gnedin} \&
  {Kaurov}}{2014}]{Gnedin2014b}
{Gnedin} N.~Y.,  {Kaurov} A.~A.,  2014, \mn@doi [\apj]
  {10.1088/0004-637X/793/1/30}, \href
  {http://adsabs.harvard.edu/abs/2014ApJ...793...30G} {793, 30}

\bibitem[\protect\citeauthoryear{{G{\'o}rski}, {Hivon}, {Banday}, {Wandelt},
  {Hansen}, {Reinecke}  \& {Bartelmann}}{{G{\'o}rski}
  et~al.}{2005}]{Gorski2005}
{G{\'o}rski} K.~M.,  {Hivon} E.,  {Banday} A.~J.,  {Wandelt} B.~D.,  {Hansen}
  F.~K.,  {Reinecke} M.,   {Bartelmann} M.,  2005, \mn@doi [\apj]
  {10.1086/427976}, \href {http://adsabs.harvard.edu/abs/2005ApJ...622..759G}
  {622, 759}

\bibitem[\protect\citeauthoryear{{Grossi} \& {Springel}}{{Grossi} \&
  {Springel}}{2009}]{Grossi2009}
{Grossi} M.,  {Springel} V.,  2009, \mn@doi [\mnras]
  {10.1111/j.1365-2966.2009.14432.x}, \href
  {http://adsabs.harvard.edu/abs/2009MNRAS.394.1559G} {394, 1559}

\bibitem[\protect\citeauthoryear{{Gunn} \& {Peterson}}{{Gunn} \&
  {Peterson}}{1965}]{Gunn1965}
{Gunn} J.~E.,  {Peterson} B.~A.,  1965, \mn@doi [\apj] {10.1086/148444}, \href
  {http://adsabs.harvard.edu/abs/1965ApJ...142.1633G} {142, 1633}

\bibitem[\protect\citeauthoryear{{Haardt} \& {Madau}}{{Haardt} \&
  {Madau}}{2012}]{Haardt2012}
{Haardt} F.,  {Madau} P.,  2012, \mn@doi [\apj] {10.1088/0004-637X/746/2/125},
  \href {http://adsabs.harvard.edu/abs/2012ApJ...746..125H} {746, 125}

\bibitem[\protect\citeauthoryear{{Hernquist} \& {Springel}}{{Hernquist} \&
  {Springel}}{2003}]{Hernquist2003}
{Hernquist} L.,  {Springel} V.,  2003, \mn@doi [\mnras]
  {10.1046/j.1365-8711.2003.06499.x}, \href
  {http://adsabs.harvard.edu/abs/2003MNRAS.341.1253H} {341, 1253}

\bibitem[\protect\citeauthoryear{Hinshaw et~al.,}{Hinshaw
  et~al.}{2013}]{Hinshaw2013}
Hinshaw G.,  et~al., 2013, The Astrophysical Journal Supplement Series, 208, 19

\bibitem[\protect\citeauthoryear{{Hockney} \& {Eastwood}}{{Hockney} \&
  {Eastwood}}{1981}]{Hockney1981}
{Hockney} R.~W.,  {Eastwood} J.~W.,  1981, {Computer Simulation Using
  Particles}

\bibitem[\protect\citeauthoryear{{Hopkins}, {Richards}  \&
  {Hernquist}}{{Hopkins} et~al.}{2007}]{Hopkins2007}
{Hopkins} P.~F.,  {Richards} G.~T.,   {Hernquist} L.,  2007, \mn@doi [\apj]
  {10.1086/509629}, \href {http://adsabs.harvard.edu/abs/2007ApJ...654..731H}
  {654, 731}

\bibitem[\protect\citeauthoryear{{Hopkins}, {Quataert}  \& {Murray}}{{Hopkins}
  et~al.}{2012}]{Hopkins2012}
{Hopkins} P.~F.,  {Quataert} E.,   {Murray} N.,  2012, \mn@doi [\mnras]
  {10.1111/j.1365-2966.2012.20593.x}, \href
  {http://adsabs.harvard.edu/abs/2012MNRAS.421.3522H} {421, 3522}

\bibitem[\protect\citeauthoryear{{Hui} \& {Gnedin}}{{Hui} \&
  {Gnedin}}{1997}]{Hui1997}
{Hui} L.,  {Gnedin} N.~Y.,  1997, \mnras, \href
  {http://adsabs.harvard.edu/abs/1997MNRAS.292...27H} {292, 27}

\bibitem[\protect\citeauthoryear{{Iliev}, {Scannapieco}  \& {Shapiro}}{{Iliev}
  et~al.}{2005}]{Iliev2005}
{Iliev} I.~T.,  {Scannapieco} E.,   {Shapiro} P.~R.,  2005, \mn@doi [\apj]
  {10.1086/429083}, \href {http://adsabs.harvard.edu/abs/2005ApJ...624..491I}
  {624, 491}

\bibitem[\protect\citeauthoryear{{Iliev}, {Mellema}, {Pen}, {Merz}, {Shapiro}
  \& {Alvarez}}{{Iliev} et~al.}{2006}]{Iliev2006}
{Iliev} I.~T.,  {Mellema} G.,  {Pen} U.-L.,  {Merz} H.,  {Shapiro} P.~R.,
  {Alvarez} M.~A.,  2006, \mn@doi [\mnras] {10.1111/j.1365-2966.2006.10502.x},
  \href {http://adsabs.harvard.edu/abs/2006MNRAS.369.1625I} {369, 1625}

\bibitem[\protect\citeauthoryear{{Iliev}, {Mellema}, {Ahn}, {Shapiro}, {Mao}
  \& {Pen}}{{Iliev} et~al.}{2014}]{Iliev2014}
{Iliev} I.~T.,  {Mellema} G.,  {Ahn} K.,  {Shapiro} P.~R.,  {Mao} Y.,   {Pen}
  U.-L.,  2014, \mn@doi [\mnras] {10.1093/mnras/stt2497}, \href
  {http://adsabs.harvard.edu/abs/2014MNRAS.439..725I} {439, 725}

\bibitem[\protect\citeauthoryear{{Jeeson-Daniel}, {Ciardi}  \&
  {Graziani}}{{Jeeson-Daniel} et~al.}{2014}]{Jeeson-Daniel2014}
{Jeeson-Daniel} A.,  {Ciardi} B.,   {Graziani} L.,  2014, \mn@doi [\mnras]
  {10.1093/mnras/stu1365}, \href
  {http://adsabs.harvard.edu/abs/2014MNRAS.443.2722J} {443, 2722}

\bibitem[\protect\citeauthoryear{{Jiang}, {Stone}  \& {Davis}}{{Jiang}
  et~al.}{2014}]{Jiang2014}
{Jiang} Y.-F.,  {Stone} J.~M.,   {Davis} S.~W.,  2014, \mn@doi [\apjs]
  {10.1088/0067-0049/213/1/7}, \href
  {http://adsabs.harvard.edu/abs/2014ApJS..213....7J} {213, 7}

\bibitem[\protect\citeauthoryear{{Kashikawa} et~al.,}{{Kashikawa}
  et~al.}{2011}]{Kashikawa2011}
{Kashikawa} N.,  et~al., 2011, \mn@doi [\apj] {10.1088/0004-637X/734/2/119},
  \href {http://adsabs.harvard.edu/abs/2011ApJ...734..119K} {734, 119}

\bibitem[\protect\citeauthoryear{{Korista}, {Baldwin}, {Ferland}  \&
  {Verner}}{{Korista} et~al.}{1997}]{Korista1997}
{Korista} K.,  {Baldwin} J.,  {Ferland} G.,   {Verner} D.,  1997, \mn@doi
  [\apjs] {10.1086/312966}, \href
  {http://adsabs.harvard.edu/abs/1997ApJS..108..401K} {108, 401}

\bibitem[\protect\citeauthoryear{{Kuhlen} \& {Faucher-Gigu{\`e}re}}{{Kuhlen} \&
  {Faucher-Gigu{\`e}re}}{2012}]{Kuhlen2012}
{Kuhlen} M.,  {Faucher-Gigu{\`e}re} C.-A.,  2012, \mn@doi [\mnras]
  {10.1111/j.1365-2966.2012.20924.x}, \href
  {http://adsabs.harvard.edu/abs/2012MNRAS.423..862K} {423, 862}

\bibitem[\protect\citeauthoryear{Leitherer et~al.,}{Leitherer
  et~al.}{1999}]{Leitherer1999}
Leitherer C.,  et~al., 1999, The Astrophysical Journal Supplement Series, 123,
  3

\bibitem[\protect\citeauthoryear{{Levermore}}{{Levermore}}{1984}]{Levermore1984}
{Levermore} C.~D.,  1984, \mn@doi [\jqsrt] {10.1016/0022-4073(84)90112-2},
  \href {http://adsabs.harvard.edu/abs/1984JQSRT..31..149L} {31, 149}

\bibitem[\protect\citeauthoryear{{Levermore} \& {Pomraning}}{{Levermore} \&
  {Pomraning}}{1981}]{Levermore1981}
{Levermore} C.~D.,  {Pomraning} G.~C.,  1981, \mn@doi [\apj] {10.1086/159157},
  \href {http://adsabs.harvard.edu/abs/1981ApJ...248..321L} {248, 321}

\bibitem[\protect\citeauthoryear{{Maselli}, {Ferrara}  \& {Ciardi}}{{Maselli}
  et~al.}{2003}]{Maselli2003}
{Maselli} A.,  {Ferrara} A.,   {Ciardi} B.,  2003, \mn@doi [\mnras]
  {10.1046/j.1365-8711.2003.06979.x}, \href
  {http://adsabs.harvard.edu/abs/2003MNRAS.345..379M} {345, 379}

\bibitem[\protect\citeauthoryear{{McQuinn}, {Lidz}, {Zahn}, {Dutta},
  {Hernquist}  \& {Zaldarriaga}}{{McQuinn} et~al.}{2007}]{McQuinn2007}
{McQuinn} M.,  {Lidz} A.,  {Zahn} O.,  {Dutta} S.,  {Hernquist} L.,
  {Zaldarriaga} M.,  2007, \mn@doi [\mnras] {10.1111/j.1365-2966.2007.11489.x},
  \href {http://adsabs.harvard.edu/abs/2007MNRAS.377.1043M} {377, 1043}

\bibitem[\protect\citeauthoryear{{McQuinn}, {Lidz}, {Zaldarriaga}, {Hernquist},
  {Hopkins}, {Dutta}  \& {Faucher-Gigu{\`e}re}}{{McQuinn}
  et~al.}{2009}]{McQuinn2009}
{McQuinn} M.,  {Lidz} A.,  {Zaldarriaga} M.,  {Hernquist} L.,  {Hopkins} P.~F.,
   {Dutta} S.,   {Faucher-Gigu{\`e}re} C.-A.,  2009, \mn@doi [\apj]
  {10.1088/0004-637X/694/2/842}, \href
  {http://adsabs.harvard.edu/abs/2009ApJ...694..842M} {694, 842}

\bibitem[\protect\citeauthoryear{{Mesinger} \& {Furlanetto}}{{Mesinger} \&
  {Furlanetto}}{2007}]{Mesinger2007}
{Mesinger} A.,  {Furlanetto} S.,  2007, \mn@doi [\apj] {10.1086/521806}, \href
  {http://adsabs.harvard.edu/abs/2007ApJ...669..663M} {669, 663}

\bibitem[\protect\citeauthoryear{{Mesinger}, {Furlanetto}  \& {Cen}}{{Mesinger}
  et~al.}{2011}]{Mesinger2011}
{Mesinger} A.,  {Furlanetto} S.,   {Cen} R.,  2011, \mn@doi [\mnras]
  {10.1111/j.1365-2966.2010.17731.x}, \href
  {http://adsabs.harvard.edu/abs/2011MNRAS.411..955M} {411, 955}

\bibitem[\protect\citeauthoryear{{Mihalas} \& {Mihalas}}{{Mihalas} \&
  {Mihalas}}{1984}]{Mihalas1984}
{Mihalas} D.,  {Mihalas} B.~W.,  1984, {Foundations of radiation hydrodynamics}

\bibitem[\protect\citeauthoryear{{Morales} \& {Wyithe}}{{Morales} \&
  {Wyithe}}{2010}]{Morales2010}
{Morales} M.~F.,  {Wyithe} J.~S.~B.,  2010, \mn@doi [\araa]
  {10.1146/annurev-astro-081309-130936}, \href
  {http://adsabs.harvard.edu/abs/2010ARA%26A..48..127M} {48, 127}

\bibitem[\protect\citeauthoryear{{Murray}, {Quataert}  \& {Thompson}}{{Murray}
  et~al.}{2010}]{Murray2010}
{Murray} N.,  {Quataert} E.,   {Thompson} T.~A.,  2010, \mn@doi [\apj]
  {10.1088/0004-637X/709/1/191}, \href
  {http://adsabs.harvard.edu/abs/2010ApJ...709..191M} {709, 191}

\bibitem[\protect\citeauthoryear{{Nakamoto}, {Umemura}  \& {Susa}}{{Nakamoto}
  et~al.}{2001}]{Nakamoto2001}
{Nakamoto} T.,  {Umemura} M.,   {Susa} H.,  2001, \mn@doi [\mnras]
  {10.1046/j.1365-8711.2001.04008.x}, \href
  {http://adsabs.harvard.edu/abs/2001MNRAS.321..593N} {321, 593}

\bibitem[\protect\citeauthoryear{{Nayakshin}, {Cha}  \& {Hobbs}}{{Nayakshin}
  et~al.}{2009}]{Nayakshin2009}
{Nayakshin} S.,  {Cha} S.-H.,   {Hobbs} A.,  2009, \mn@doi [\mnras]
  {10.1111/j.1365-2966.2009.15091.x}, \href
  {http://adsabs.harvard.edu/abs/2009MNRAS.397.1314N} {397, 1314}

\bibitem[\protect\citeauthoryear{{Norman}, {Reynolds}, {So}, {Harkness}  \&
  {Wise}}{{Norman} et~al.}{2015}]{Norman2013}
{Norman} M.~L.,  {Reynolds} D.~R.,  {So} G.~C.,  {Harkness} R.~P.,   {Wise}
  J.~H.,  2015, \mn@doi [\apjs] {10.1088/0067-0049/216/1/16}, \href
  {http://adsabs.harvard.edu/abs/2015ApJS..216...16N} {216, 16}

\bibitem[\protect\citeauthoryear{{Oesch} et~al.,}{{Oesch}
  et~al.}{2014}]{Oesch2014}
{Oesch} P.~A.,  et~al., 2014, \mn@doi [\apj] {10.1088/0004-637X/786/2/108},
  \href {http://adsabs.harvard.edu/abs/2014ApJ...786..108O} {786, 108}

\bibitem[\protect\citeauthoryear{{Ouchi} et~al.,}{{Ouchi}
  et~al.}{2010}]{Ouchi2010}
{Ouchi} M.,  et~al., 2010, \mn@doi [\apj] {10.1088/0004-637X/723/1/869}, \href
  {http://adsabs.harvard.edu/abs/2010ApJ...723..869O} {723, 869}

\bibitem[\protect\citeauthoryear{{Paardekooper}, {Khochfar}  \& {Dalla
  Vecchia}}{{Paardekooper} et~al.}{2013}]{Pardekooper2013}
{Paardekooper} J.-P.,  {Khochfar} S.,   {Dalla Vecchia} C.,  2013, \mn@doi
  [\mnras] {10.1093/mnrasl/sls032}, \href
  {http://adsabs.harvard.edu/abs/2013MNRAS.429L..94P} {429, L94}

\bibitem[\protect\citeauthoryear{{Parsons} et~al.,}{{Parsons}
  et~al.}{2014}]{Parsons2014}
{Parsons} A.~R.,  et~al., 2014, \mn@doi [\apj] {10.1088/0004-637X/788/2/106},
  \href {http://adsabs.harvard.edu/abs/2014ApJ...788..106P} {788, 106}

\bibitem[\protect\citeauthoryear{{Pawlik} \& {Schaye}}{{Pawlik} \&
  {Schaye}}{2008}]{Pawlik2008}
{Pawlik} A.~H.,  {Schaye} J.,  2008, \mn@doi [\mnras]
  {10.1111/j.1365-2966.2008.13601.x}, \href
  {http://adsabs.harvard.edu/abs/2008MNRAS.389..651P} {389, 651}

\bibitem[\protect\citeauthoryear{{Pawlik}, {Schaye}  \& {van
  Scherpenzeel}}{{Pawlik} et~al.}{2009}]{Pawlik2009}
{Pawlik} A.~H.,  {Schaye} J.,   {van Scherpenzeel} E.,  2009, \mn@doi [\mnras]
  {10.1111/j.1365-2966.2009.14486.x}, \href
  {http://adsabs.harvard.edu/abs/2009MNRAS.394.1812P} {394, 1812}

\bibitem[\protect\citeauthoryear{{Pawlik}, {Schaye}  \& {Dalla
  Vecchia}}{{Pawlik} et~al.}{2015}]{Pawlik2015}
{Pawlik} A.~H.,  {Schaye} J.,   {Dalla Vecchia} C.,  2015, \mn@doi [\mnras]
  {10.1093/mnras/stv976}, \href
  {http://adsabs.harvard.edu/abs/2015MNRAS.451.1586P} {451, 1586}

\bibitem[\protect\citeauthoryear{{Pentericci} et~al.,}{{Pentericci}
  et~al.}{2011}]{Pentericci2011}
{Pentericci} L.,  et~al., 2011, \mn@doi [\apj] {10.1088/0004-637X/743/2/132},
  \href {http://adsabs.harvard.edu/abs/2011ApJ...743..132P} {743, 132}

\bibitem[\protect\citeauthoryear{{Petkova} \& {Springel}}{{Petkova} \&
  {Springel}}{2009}]{Petkova2009}
{Petkova} M.,  {Springel} V.,  2009, \mn@doi [\mnras]
  {10.1111/j.1365-2966.2009.14843.x}, \href
  {http://adsabs.harvard.edu/abs/2009MNRAS.396.1383P} {396, 1383}

\bibitem[\protect\citeauthoryear{{Petkova} \& {Springel}}{{Petkova} \&
  {Springel}}{2011a}]{Petkova2011b}
{Petkova} M.,  {Springel} V.,  2011a, \mn@doi [\mnras]
  {10.1111/j.1365-2966.2010.17955.x}, \href
  {http://adsabs.harvard.edu/abs/2011MNRAS.412..935P} {412, 935}

\bibitem[\protect\citeauthoryear{{Petkova} \& {Springel}}{{Petkova} \&
  {Springel}}{2011b}]{Petkova2011}
{Petkova} M.,  {Springel} V.,  2011b, \mn@doi [\mnras]
  {10.1111/j.1365-2966.2011.18986.x}, \href
  {http://adsabs.harvard.edu/abs/2011MNRAS.415.3731P} {415, 3731}

\bibitem[\protect\citeauthoryear{{Pillepich} et~al.,}{{Pillepich}
  et~al.}{2014}]{Pillepich2014}
{Pillepich} A.,  et~al., 2014, \mn@doi [\mnras] {10.1093/mnras/stu1408}, \href
  {http://adsabs.harvard.edu/abs/2014MNRAS.444..237P} {444, 237}

\bibitem[\protect\citeauthoryear{{Planck Collaboration} et~al.,}{{Planck
  Collaboration} et~al.}{2014}]{Planck2013CosmoParams}
{Planck Collaboration} et~al., 2014, \mn@doi [\aap]
  {10.1051/0004-6361/201321591}, \href
  {http://adsabs.harvard.edu/abs/2014A%26A...571A..16P} {571, A16}

\bibitem[\protect\citeauthoryear{{Planck Collaboration} et~al.,}{{Planck
  Collaboration} et~al.}{2015}]{PlanckCollaboration2015}
{Planck Collaboration} et~al., 2015, preprint, \href
  {http://adsabs.harvard.edu/abs/2015arXiv150201582P} {} (\mn@eprint {arXiv}
  {1502.01582})

\bibitem[\protect\citeauthoryear{{Rahmati}, {Pawlik}, {Rai{\v c}evi{\'c}}  \&
  {Schaye}}{{Rahmati} et~al.}{2013}]{Rahmati2013}
{Rahmati} A.,  {Pawlik} A.~H.,  {Rai{\v c}evi{\'c}} M.,   {Schaye} J.,  2013,
  \mn@doi [\mnras] {10.1093/mnras/stt066}, \href
  {http://adsabs.harvard.edu/abs/2013MNRAS.430.2427R} {430, 2427}

\bibitem[\protect\citeauthoryear{{Rai{\v c}evi{\'c}}, {Theuns}  \&
  {Lacey}}{{Rai{\v c}evi{\'c}} et~al.}{2011}]{Raicevic2011}
{Rai{\v c}evi{\'c}} M.,  {Theuns} T.,   {Lacey} C.,  2011, \mn@doi [\mnras]
  {10.1111/j.1365-2966.2010.17480.x}, \href
  {http://adsabs.harvard.edu/abs/2011MNRAS.410..775R} {410, 775}

\bibitem[\protect\citeauthoryear{{Razoumov}, {Norman}, {Abel}  \&
  {Scott}}{{Razoumov} et~al.}{2002}]{Razoumov2002}
{Razoumov} A.~O.,  {Norman} M.~L.,  {Abel} T.,   {Scott} D.,  2002, \mn@doi
  [\apj] {10.1086/340451}, \href
  {http://adsabs.harvard.edu/abs/2002ApJ...572..695R} {572, 695}

\bibitem[\protect\citeauthoryear{{Robertson}, {Ellis}, {Furlanetto}  \&
  {Dunlop}}{{Robertson} et~al.}{2015}]{Robertson2015}
{Robertson} B.~E.,  {Ellis} R.~S.,  {Furlanetto} S.~R.,   {Dunlop} J.~S.,
  2015, \mn@doi [\apjl] {10.1088/2041-8205/802/2/L19}, \href
  {http://adsabs.harvard.edu/abs/2015ApJ...802L..19R} {802, L19}

\bibitem[\protect\citeauthoryear{{Rodriguez-Gomez} et~al.,}{{Rodriguez-Gomez}
  et~al.}{2015}]{Rodriguez-Gomez2015}
{Rodriguez-Gomez} V.,  et~al., 2015, \mn@doi [\mnras] {10.1093/mnras/stv264},
  \href {http://adsabs.harvard.edu/abs/2015MNRAS.449...49R} {449, 49}

\bibitem[\protect\citeauthoryear{{Rosdahl}, {Blaizot}, {Aubert}, {Stranex}  \&
  {Teyssier}}{{Rosdahl} et~al.}{2013}]{Rosdahl2013}
{Rosdahl} J.,  {Blaizot} J.,  {Aubert} D.,  {Stranex} T.,   {Teyssier} R.,
  2013, \mn@doi [\mnras] {10.1093/mnras/stt1722}, \href
  {http://adsabs.harvard.edu/abs/2013MNRAS.436.2188R} {436, 2188}

\bibitem[\protect\citeauthoryear{{Rosdahl}, {Schaye}, {Teyssier}  \&
  {Agertz}}{{Rosdahl} et~al.}{2015}]{Rosdahl2015}
{Rosdahl} J.,  {Schaye} J.,  {Teyssier} R.,   {Agertz} O.,  2015, preprint,
  \href {http://adsabs.harvard.edu/abs/2015arXiv150104632R} {} (\mn@eprint
  {arXiv} {1501.04632})

\bibitem[\protect\citeauthoryear{{Sales}, {Marinacci}, {Springel}  \&
  {Petkova}}{{Sales} et~al.}{2014}]{Sales2014}
{Sales} L.~V.,  {Marinacci} F.,  {Springel} V.,   {Petkova} M.,  2014, \mn@doi
  [\mnras] {10.1093/mnras/stu155}, \href
  {http://adsabs.harvard.edu/abs/2014MNRAS.439.2990S} {439, 2990}

\bibitem[\protect\citeauthoryear{{Sales} et~al.,}{{Sales}
  et~al.}{2015}]{Sales2015}
{Sales} L.~V.,  et~al., 2015, \mn@doi [\mnras] {10.1093/mnrasl/slu173}, \href
  {http://adsabs.harvard.edu/abs/2015MNRAS.447L...6S} {447, L6}

\bibitem[\protect\citeauthoryear{{Scannapieco} et~al.,}{{Scannapieco}
  et~al.}{2012}]{Scannapieco2012}
{Scannapieco} C.,  et~al., 2012, \mn@doi [\mnras]
  {10.1111/j.1365-2966.2012.20993.x}, \href
  {http://adsabs.harvard.edu/abs/2012MNRAS.423.1726S} {423, 1726}

\bibitem[\protect\citeauthoryear{{Schaye}, {Theuns}, {Rauch}, {Efstathiou}  \&
  {Sargent}}{{Schaye} et~al.}{2000}]{Schaye2000}
{Schaye} J.,  {Theuns} T.,  {Rauch} M.,  {Efstathiou} G.,   {Sargent} W.~L.~W.,
   2000, \mn@doi [\mnras] {10.1046/j.1365-8711.2000.03815.x}, \href
  {http://adsabs.harvard.edu/abs/2000MNRAS.318..817S} {318, 817}

\bibitem[\protect\citeauthoryear{{Schaye} et~al.,}{{Schaye}
  et~al.}{2015}]{Schaye2014}
{Schaye} J.,  et~al., 2015, \mn@doi [\mnras] {10.1093/mnras/stu2058}, \href
  {http://adsabs.harvard.edu/abs/2015MNRAS.446..521S} {446, 521}

\bibitem[\protect\citeauthoryear{{Sijacki}, {Vogelsberger}, {Genel},
  {Springel}, {Torrey}, {Snyder}, {Nelson}  \& {Hernquist}}{{Sijacki}
  et~al.}{2015}]{Sijacki2014}
{Sijacki} D.,  {Vogelsberger} M.,  {Genel} S.,  {Springel} V.,  {Torrey} P.,
  {Snyder} G.~F.,  {Nelson} D.,   {Hernquist} L.,  2015, \mn@doi [\mnras]
  {10.1093/mnras/stv1340}, \href
  {http://adsabs.harvard.edu/abs/2015MNRAS.452..575S} {452, 575}

\bibitem[\protect\citeauthoryear{{So}, {Norman}, {Reynolds}  \& {Wise}}{{So}
  et~al.}{2014}]{So2014}
{So} G.~C.,  {Norman} M.~L.,  {Reynolds} D.~R.,   {Wise} J.~H.,  2014, \mn@doi
  [\apj] {10.1088/0004-637X/789/2/149}, \href
  {http://adsabs.harvard.edu/abs/2014ApJ...789..149S} {789, 149}

\bibitem[\protect\citeauthoryear{{Sokasian}, {Abel}  \& {Hernquist}}{{Sokasian}
  et~al.}{2001}]{Sokasian2001}
{Sokasian} A.,  {Abel} T.,   {Hernquist} L.~E.,  2001, \mn@doi [\na]
  {10.1016/S1384-1076(01)00065-3}, \href
  {http://adsabs.harvard.edu/abs/2001NewA....6..359S} {6, 359}

\bibitem[\protect\citeauthoryear{{Springel}}{{Springel}}{2010}]{Springel2010}
{Springel} V.,  2010, \mn@doi [\mnras] {10.1111/j.1365-2966.2009.15715.x},
  \href {http://adsabs.harvard.edu/abs/2010MNRAS.401..791S} {401, 791}

\bibitem[\protect\citeauthoryear{{Springel}, {White}, {Tormen}  \&
  {Kauffmann}}{{Springel} et~al.}{2001}]{Springel2001}
{Springel} V.,  {White} S.~D.~M.,  {Tormen} G.,   {Kauffmann} G.,  2001,
  \mn@doi [\mnras] {10.1046/j.1365-8711.2001.04912.x}, \href
  {http://adsabs.harvard.edu/abs/2001MNRAS.328..726S} {328, 726}

\bibitem[\protect\citeauthoryear{{Springel} et~al.,}{{Springel}
  et~al.}{2008}]{Springel2008}
{Springel} V.,  et~al., 2008, \mn@doi [\mnras]
  {10.1111/j.1365-2966.2008.14066.x}, \href
  {http://adsabs.harvard.edu/abs/2008MNRAS.391.1685S} {391, 1685}

\bibitem[\protect\citeauthoryear{{Stinson}, {Brook}, {Macci{\`o}}, {Wadsley},
  {Quinn}  \& {Couchman}}{{Stinson} et~al.}{2013}]{Stinson2013}
{Stinson} G.~S.,  {Brook} C.,  {Macci{\`o}} A.~V.,  {Wadsley} J.,  {Quinn}
  T.~R.,   {Couchman} H.~M.~P.,  2013, \mn@doi [\mnras] {10.1093/mnras/sts028},
  \href {http://adsabs.harvard.edu/abs/2013MNRAS.428..129S} {428, 129}

\bibitem[\protect\citeauthoryear{{Torrey}, {Vogelsberger}, {Genel}, {Sijacki},
  {Springel}  \& {Hernquist}}{{Torrey} et~al.}{2014}]{Torrey2014}
{Torrey} P.,  {Vogelsberger} M.,  {Genel} S.,  {Sijacki} D.,  {Springel} V.,
  {Hernquist} L.,  2014, \mn@doi [\mnras] {10.1093/mnras/stt2295}, \href
  {http://adsabs.harvard.edu/abs/2014MNRAS.438.1985T} {438, 1985}

\bibitem[\protect\citeauthoryear{{Torrey} et~al.,}{{Torrey}
  et~al.}{2015}]{Torrey2015}
{Torrey} P.,  et~al., 2015, \mn@doi [\mnras] {10.1093/mnras/stu2592}, \href
  {http://adsabs.harvard.edu/abs/2015MNRAS.447.2753T} {447, 2753}

\bibitem[\protect\citeauthoryear{{Trac} \& {Cen}}{{Trac} \&
  {Cen}}{2007}]{Trac2007}
{Trac} H.,  {Cen} R.,  2007, \mn@doi [\apj] {10.1086/522566}, \href
  {http://adsabs.harvard.edu/abs/2007ApJ...671....1T} {671, 1}

\bibitem[\protect\citeauthoryear{{Trac}, {Cen}  \& {Loeb}}{{Trac}
  et~al.}{2008}]{Trac2008}
{Trac} H.,  {Cen} R.,   {Loeb} A.,  2008, \mn@doi [\apjl] {10.1086/595678},
  \href {http://adsabs.harvard.edu/abs/2008ApJ...689L..81T} {689, L81}

\bibitem[\protect\citeauthoryear{{Turner} \& {Stone}}{{Turner} \&
  {Stone}}{2001}]{Turner2001}
{Turner} N.~J.,  {Stone} J.~M.,  2001, \mn@doi [\apjs] {10.1086/321779}, \href
  {http://adsabs.harvard.edu/abs/2001ApJS..135...95T} {135, 95}

\bibitem[\protect\citeauthoryear{{Vogelsberger}, {Genel}, {Sijacki}, {Torrey},
  {Springel}  \& {Hernquist}}{{Vogelsberger} et~al.}{2013}]{Vogelsberger2013}
{Vogelsberger} M.,  {Genel} S.,  {Sijacki} D.,  {Torrey} P.,  {Springel} V.,
  {Hernquist} L.,  2013, \mn@doi [\mnras] {10.1093/mnras/stt1789}, \href
  {http://adsabs.harvard.edu/abs/2013MNRAS.436.3031V} {436, 3031}

\bibitem[\protect\citeauthoryear{{Vogelsberger} et~al.,}{{Vogelsberger}
  et~al.}{2014a}]{Vogelsberger2014b}
{Vogelsberger} M.,  et~al., 2014a, \mn@doi [\mnras] {10.1093/mnras/stu1536},
  \href {http://adsabs.harvard.edu/abs/2014MNRAS.444.1518V} {444, 1518}

\bibitem[\protect\citeauthoryear{{Vogelsberger} et~al.,}{{Vogelsberger}
  et~al.}{2014b}]{Vogelsberger2014}
{Vogelsberger} M.,  et~al., 2014b, \mn@doi [\nat] {10.1038/nature13316}, \href
  {http://adsabs.harvard.edu/abs/2014Natur.509..177V} {509, 177}

\bibitem[\protect\citeauthoryear{{Wellons} et~al.,}{{Wellons}
  et~al.}{2015}]{Wellons2014}
{Wellons} S.,  et~al., 2015, \mn@doi [\mnras] {10.1093/mnras/stv303}, \href
  {http://adsabs.harvard.edu/abs/2015MNRAS.449..361W} {449, 361}

\bibitem[\protect\citeauthoryear{{Wetterich}}{{Wetterich}}{2004}]{Wetterich2004}
{Wetterich} C.,  2004, \mn@doi [Physics Letters B]
  {10.1016/j.physletb.2004.05.008}, \href
  {http://adsabs.harvard.edu/abs/2004PhLB..594...17W} {594, 17}

\bibitem[\protect\citeauthoryear{{Wise}, {Demchenko}, {Halicek}, {Norman},
  {Turk}, {Abel}  \& {Smith}}{{Wise} et~al.}{2014}]{Wise2014}
{Wise} J.~H.,  {Demchenko} V.~G.,  {Halicek} M.~T.,  {Norman} M.~L.,  {Turk}
  M.~J.,  {Abel} T.,   {Smith} B.~D.,  2014, \mn@doi [\mnras]
  {10.1093/mnras/stu979}, \href
  {http://adsabs.harvard.edu/abs/2014MNRAS.442.2560W} {442, 2560}

\bibitem[\protect\citeauthoryear{{Yoshida}, {Sokasian}, {Hernquist}  \&
  {Springel}}{{Yoshida} et~al.}{2003a}]{Yoshida2003a}
{Yoshida} N.,  {Sokasian} A.,  {Hernquist} L.,   {Springel} V.,  2003a, \mn@doi
  [\apjl] {10.1086/376963}, \href
  {http://adsabs.harvard.edu/abs/2003ApJ...591L...1Y} {591, L1}

\bibitem[\protect\citeauthoryear{{Yoshida}, {Sokasian}, {Hernquist}  \&
  {Springel}}{{Yoshida} et~al.}{2003b}]{Yoshida2003b}
{Yoshida} N.,  {Sokasian} A.,  {Hernquist} L.,   {Springel} V.,  2003b, \mn@doi
  [\apj] {10.1086/378852}, \href
  {http://adsabs.harvard.edu/abs/2003ApJ...598...73Y} {598, 73}

\bibitem[\protect\citeauthoryear{{Zahn}, {Lidz}, {McQuinn}, {Dutta},
  {Hernquist}, {Zaldarriaga}  \& {Furlanetto}}{{Zahn} et~al.}{2007}]{Zahn2007}
{Zahn} O.,  {Lidz} A.,  {McQuinn} M.,  {Dutta} S.,  {Hernquist} L.,
  {Zaldarriaga} M.,   {Furlanetto} S.~R.,  2007, \mn@doi [\apj]
  {10.1086/509597}, \href {http://adsabs.harvard.edu/abs/2007ApJ...654...12Z}
  {654, 12}

\bibitem[\protect\citeauthoryear{{Zahn}, {Mesinger}, {McQuinn}, {Trac}, {Cen}
  \& {Hernquist}}{{Zahn} et~al.}{2011}]{Zahn2011}
{Zahn} O.,  {Mesinger} A.,  {McQuinn} M.,  {Trac} H.,  {Cen} R.,   {Hernquist}
  L.~E.,  2011, \mn@doi [\mnras] {10.1111/j.1365-2966.2011.18439.x}, \href
  {http://adsabs.harvard.edu/abs/2011MNRAS.414..727Z} {414, 727}

\bibitem[\protect\citeauthoryear{{Zaroubi}}{{Zaroubi}}{2013}]{Zaroubi2013}
{Zaroubi} S.,  2013, in {Wiklind} T.,  {Mobasher} B.,   {Bromm} V.,  eds,
  Astrophysics and Space Science Library Vol. 396, Astrophysics and Space
  Science Library. p.~45 (\mn@eprint {arXiv} {1206.0267})

\makeatother
\end{thebibliography}

\bsp	
\label{lastpage}
\end{document}